\documentclass[preprint,12pt]{article}

\usepackage{lineno,hyperref,amsmath,amssymb,color}
\usepackage{graphics} 
\usepackage{epsfig}
\usepackage{graphicx} % gr‡ficos
\usepackage{subfigure} % subgr‡ficos
\usepackage{graphicx}
\usepackage{epstopdf}
\modulolinenumbers[5]

\usepackage{epstopdf}% To incorporate .eps illustrations using PDFLaTeX, etc.
\usepackage[colorinlistoftodos]{todonotes}
\usepackage{graphicx} % graficos
\usepackage{psfrag} 
\usepackage{amsthm}
\usepackage{amsmath, amssymb, dsfont}
\usepackage{bm}
\usepackage{ulem}
\usepackage{geometry}
\newtheorem{theorem}{Theorem}[section]

\newtheorem{lemma}[theorem]{Lemma}

\newcommand\redout{\bgroup\markoverwith{\textcolor{red}{\rule[.5ex]{2pt}{0.8pt}}}\ULon}
\usepackage{color}

\begin{document}
%\begin{frontmatter}
\title{Predictive data assimilation through Reduced Order Modeling for epidemics with data uncertainty}
\author{T. Chac\'on Rebollo\footnotemark[1]
\and D. Franco Coronil\footnotemark[2]}
\footnotetext[1]
{Dpto. EDAN \& IMUS, University of Seville, Campus de Reina Mercedes, 41012 Sevilla (Spain), e-mail: chacon@us.es}
\footnotetext[2]
{Dpto. EDAN, University of Seville, Campus de Reina Mercedes, 41012 Sevilla (Spain), e-mail: franco@us.es}
%\author{T. Chac\'on Rebollo,\footnote{Corresponding author: chacon@us.es}}
%\address{Dpto. EDAN\&IMUS, University of Seville, \\ Campus de Reina Mercedes, 41012 Sevilla, Spain}
%\author{D. Franco Coronil}
%\address{Dpto. EDAN, University of Seville, \\ Campus de Reina Mercedes, 41012 Sevilla, Spain}
\maketitle
\begin{abstract}
In this article, we develop a data assimilation procedure to predict the evolution of epidemics with data uncertainty, with application to the Covid-19 pandemic. We construct a vademecum of solutions by solving the SIR epidemic model for a set of data neighboring the estimated real (or official) ones. A reduced basis is constructed from this vademecum through Proper Orthogonal Decomposition (POD). The reduced POD base is then applied to assimilate the pandemic data (infected, recovered, deceased) during the period in which data are known, by a least squares procedure. The fitted curves are then used to predict the evolution of the pandemic in the next days. Validation tests for Andalusia region (Spain), Italy and Spain show accurate predictions for 7 days that improve as the number of assimilated data increases.
\end{abstract}


%\end{frontmatter}

\section{Introduction} \label{sec:intro}
One of the largest difficulties when dealing with predictive estimates of the evolution of the Covid-19 pandemic is the lack of reliable data on real number of infected people, as well as infection, recovery and death rates. There are, however, reference data that present some variability from a country or region to another.  Moreover these parameters in each actual country or region may vary in time due to the relaxation of the population in the respect of the lockdown measures or changes in the availability of health care resources, among other factors. Moreover, there is a number of reliable data, essentially the recovered and deceased populations.

The use of standard mathematical models of the epidemic thus faces important incertitudes. These may be afforded with statistical techniques, that should be based upon medical tests, and are scarce in some countries. Also with least squares techniques, that allow to recover estimates for the lacking parameters.

We afford here the use of Reduced Order Modeling (ROM) to approximate the evolution of the pandemic and predict its future evolution. ROM allows to extract the dominant patterns of parametric systems, providing approximating spaces of very low dimension and excellent accuracy properties. 

Our approach is based upon the construction of a vademecum of solutions if the SIR model, for values of the parameters neighboring the reference ones and long time periods of evolution of the pandemic. A reduced basis is then constructed by the Proper Orthogonal Decompostion (POD), that approximates with good accuracy all the solutions in the vademecum. Least squares is then used to approximate on this reduced space the reliable data for a time period in which these are known. This circumvents the need of accurate values for the parameters to obtain evolution curves of the pandemic. The fitted curves are then used to estimate the future evolution of the pandemic.

We perform some validation tests with data of the evolution of the pandemic with analytic functions and solutions of the SIR model,  that show that the procedure provides quite accurate fitting if the number of data is large enough. Also that assimilating data in a small time interval to a very small number of POD modes provides a qualitative good approximation of the evolution of the epidemic with relatively few elements in the vademecum. This approximation becomes fairly accurate if the vademecum is conveniently enriched. We apply the procedure to predict the evolution of the Covid-19 pandemic in Andalusia region, Spain and Italy. In despite of the fact that the official data on active infected people are well known to be far from the real ones, we obtain good accuracy in the prediction of the pandemic evolution in the next 7 days, with relative errors in infected, recovered and deceased populations typically below 4\%, that are smaller for larger numbers of assimilated data increase.

The paper is organized as follows. The construction of the vademecum of solution is described in Section \ref{se:vademecum}, the assimilation method is described in Section \ref{se:rb} and the least squares procedure in Section \ref{se:lsq}. Section \ref{se:validation} deals with the validation of the procedure, while finally Section \ref{se:conclusions} presents some conclusions.

\section{Vademecum of model solutions}\label{se:vademecum}
 We start form the basic SIR model, with variable infection rate, that we assume to correctly describe the pandemic:
 \begin{equation} \label{modsir}
 \left \{ \begin{array}{rcl}
 S'&=&-\alpha(t)\, \displaystyle\frac{S}{S_\infty}\, I , \\
 I'&=& \,\,\,\,\displaystyle\alpha(t)\, \frac{S}{S_\infty}\, I - \beta \, I, \\
 R'&=&\beta \, I.
 \end{array}
 \right .
 \end{equation}
 Here $S$, $I$, and $R$ respectively are the number of susceptible, infected and recovered individuals; $S_\infty$ is the total population, $\alpha$ is the transmission rate, that we assume variable in time, and $\beta$ is the recovery rate. The function $\alpha(t)$ and the parameter $\beta$ respectively are the inverse of the characteristic infection $T_i(t)$ and recovery times $T_r$. If we consider the basic reproduction rate
 $$
 r_0(t)=\frac{\alpha(t)}{\beta}=\frac{T_r}{T_i(t)},
 $$
 we write the model as 
\begin{equation} \label{modsir2}
 \left \{ \begin{array}{rcl}
 S'&=&-\beta \, R_0(t)\, \displaystyle\frac{S}{S_\infty}\, I , \\
 I'&=& \,\,\,\,\displaystyle\beta \, \left (r_0(t)\, \frac{S}{S_\infty}\, -  1\right )\, I, \\
 R'&=&\beta \, I.
 \end{array}
 \right .
 \end{equation}
 This model should be complemented with initial conditions 
 \begin{equation} \label{modsir2i} I(0)=I_0, \quad R(0)=R_0,\quad S(0)=S_0=S_\infty-I_0-R_0.
 \end{equation}
 \par
 In practice we shall approximate $r_0$ by piecewise constant functions, taking into consideration the relaxation of the lockdown restriction as times goes on.
 
 We know reference values for $r_0$ for the Covid-19 epidemic, ranging from 0.5 when the lockdown measures take place for time enough, and up to 3 if no protection measures are taken. Also, a reference values for the recovery time $T_r$ is 15 (cf. Guti\'errez and Varona \cite{Varona}, Jiwei et al. \cite{chinos}). 
 
 The real amount of infected people $I_0$ is far from being known, there are several estimates that give factors of around 15 times the official values. For instance in France the value of initially infected possibly is nearly 14 times the official one (cf. Roque et al. \cite{franceses}). However we shall take as reference values for $I_0$, and for $R_0$, the official number of infected and recovered people, respectively, at the initial time of our computation. As we shall see it will not be necessary to provide accurate approximations of the initial values of infected population to obtain good approximations of the recovered and deceased populations, which are the ones that affect the health care systems.
 
 To construct our vademecum of solutions we define $r_0$ as
 \begin{equation}\label{r0variable}
 r_0(t)= \left \{ \begin{array}{ccc}
 r_{ref,1} &\mbox{if } & 0< t < T_1,\\
 r_{ref,2} &\mbox{if } & T_1\le t < T,\\
 \end{array}
 \right .
 \end{equation}
 where $r_{ref,1} $ and $r_{ref,2} $ are reference values for $r_0$ in the time intervals $[0,T_1]$ and $[T_1,T]$. The first time interval corresponds to the initial phase of the lockdown with strict fitting of protection measures by the people (21 days). The second one correspond to a second phase in which there is some relaxation of the lockdown measures. We have taken $r_{ref,1}= 1.1 $ and $r_{ref,2}=0.8 $.
 
 We do not consider the initial phase of the epidemic in which no protection measures are taken, this period can be skipped as the data during the time interval $[0,T]$ are known. Our purpose is to use these data to predict the evolution of the epidemic in times later than $T$.
 
 We build the vademecum of solutions by solving the SIR model \eqref{modsir2} for a set of discrete values of the parameters $r_{ref,1}$, $r_{ref,2}$, $T_r$, $I_0$ and $R_0$ neighboring the reference values. We typically take 4 for both the initial conditions and for the parameters, equally spaced. The computation time $T$ is large enough to let the epidemic reach its steady state for all set of parameters considered. We use the ODE45 Mathlab solver to perform the computations, with a time step of 0.01 days.
 
 We denote the vademecum of solutions as $\{I_p\}_{p=1}^P$, $\{R_p\}_{p=1}^P$, where $P$ is the total number of runs generated by combining the different discrete values of the parameters. We assume that we know $I_p$ and $R_p$ at discrete times $t_n = n\,\Delta t$, $n=0,1,\cdots,M$, so that $T=M \,\Delta t$, and denote $I_p^n=I_p(t_n)$ and similarly $R_p^n$. In practice we take $\Delta t= 1$ day.

\section{Construction of Reduced Basis}\label{se:rb}
To build the reduced basis we apply the Proper Orthogonal Decomposition (POD) to the vademecum of solutions computed as described in the preceding section. Our claim is that if the true parameters that govern the epidemic lie within the range of parameters that we consider to build the vademecum, then the essential patterns of the epidemic will be well approximated by the linear space spanned by the vademecum solutions.  Indeed, the solution of the initial value problem \eqref{modsir2}-\eqref{modsir2i} depends continuously on the initial data, parameter $\beta$ and reproduction rate $r_0(t)$. If we denote its solution by $F_{F_0,\beta, r_0}(t)$ with $F_{F_0,\beta, r_0}(t)=(S(t),I(t),R(t)) \in \mathbb{R}^3$, $F_0=(S_0,I_0,R_0) \in \mathbb{R}^3$, then it holds
\begin{lemma} Assume that $S_0 \ge 0$, $I_0 \ge 0$, $R_0 \ge 0$,  that $\beta$, $\beta'$ lie in a bounded set ${\cal B} \subset \mathbb{R}$ and that $r_0$, $r_0'$ lie in a bounded set ${\cal R} \subset L^\infty(0,T)$. Then
$$
 \|F_{F_0,\beta, r_0}(t)- F_{F_0',\beta', r_0'}(t)\|_{\mathbb{R}^3} \le \|F_0-F'_0\|_{\mathbb{R}^3}\, e^{a\,t} + \frac{b}{a}\, \left (e^{a\,t}-1\right )\quad \forall t \in [0,T],
$$
where 
\begin{eqnarray*} 
&&a= (2+  (S_\infty +1)\,R )\,B ,\quad b= 2 
 (S_\infty +1)\,R\, |\beta-\beta'|+S_\infty \,B\, \|r_0-r_0'\|_{ L^\infty(0,T)} ,
\end{eqnarray*}
with
$$B= \max_{\beta \in {\cal B}}|\beta|,\quad R=\max_{r_0 \in {\cal R}}\|r_0\|_{ L^\infty(0,T)}.
$$
\end{lemma}
{\bf Proof:} The proof is standard by Gronwall's Lemma, using that $S$, $I$ and $R$ are non-negative and bounded by $S_\infty$. \hfill $\Box$.

Then if our discretization of the parameters space is fine enough, we will obtain a good approximation of the epidemic evolution in the space spanned by the solution of the SIR model.

As a consequence, in principle a least-squares fitting of the true epidemic data on the space spanned by the parametric solutions of the SIR model could give a good approximation of the evolution of the epidemic. However the grassmaniann matrices associated to the vademecum are singular up to computer precision.

 Instead we consider a reduced order approximation of the parametric solutions space. The POD allows to construct a reduced basis that retains the dominant patterns in the vademecum, and generate a space that provides a good approximation of the one spanned by the vademecum functions.

The Proper Orthogonal (or Karhunen-Lo\`eve) decomposition  provides a technique to obtain low-dimensional approximations of parametric functions. To describe it in our framework (cf. Aza\"{\i}ez et al.\cite{RPOD}), let us consider a Hilbert space $H$ of finite dimension, endowed with a scalar product $(\cdot,\cdot)_H$, and a parameter measure space $G$ endowed with a measure $\mu$. In our case $H=\mathbb{R}^K$ (for some $K$ that we shall specify in the sequel) endowed with the discrete $l^2(0,t_K)$ inner product,
$$
(u,w)_H=\Delta t \, \sum_{m=0}^K u_m\, w_m,\quad \forall   u, \, w \in H;
$$
and $G=\{1,2,\cdots,P\}$ endowed with the discrete measure given by
$$
\mu(S)=\mbox{card}(S),\quad \forall S \subset G,
$$
where card($S$) denotes the cardinal of the set $S$. Let us consider a function $f \in L^2(G, H;d\mu)$, and introduce the POD operator
$$
A: H \mapsto H, \quad \mbox A\psi=  \int_Gf(\gamma)\, (f(\gamma),\psi)_H\, d\mu(\gamma) =\sum_{p=1}^P f_p\, \Delta t\, \sum_{i=1}^K f_p^i \,\psi_i \quad \mbox{for  } \psi \in H ,
$$
where we denote $f_p=f(p)$ and $f_p=(f_p^1,\cdots, f_p^K)\in H$, for all $p\in G$. The POD operator is trivially linear and bounded. Moreover it is  self-adjoint and non-negative, as $A=B^*B$, where $B:H \mapsto L^2(G;d\mu)$ and its adjoint operator $B^*:L^2(G;d\mu) \mapsto H$ are given by (cf. Muller \cite{muller}, Chapter 2)
\begin{equation} \label{eq:opbbt}
(B\varphi)(\gamma)=(f(\gamma),\varphi)_H \quad\mbox{for  } \varphi \in H, \quad
B^* v =  (v,f)_{L^2(G;d\mu)} \quad\mbox{for  } v \in L^2(G;d\mu).
\end{equation}
 Consequently, there exists an orthonormal basis of $H$ formed by eigenvectors $\{v^m\}_{m \ge 0}$ of $A$, associated to non-negative eigenvalues $\{\lambda_m\}_{m \ge 0}$. 

The main interest of the POD is the following best-approximation property (cf. \cite{muller}, Chapter 2):
\begin{lemma} \label{le:bestapp}
Let  $V_l=Span(v^1,\cdots,v^l)\subset H$. Let $W_l$ be any sub-space of $H$ of dimension $l$. Then
$$
\int_G d_H(f(\gamma),V_l)^2\, d\gamma \le \int_G d_H(f(\gamma),W_l)^2\, d\gamma,
$$
where
$$
d_H(\varphi,W_l)=\inf_{\psi \in W_l} \|\varphi-\psi\|_H\quad \mbox{for  } \varphi \in H
$$
denotes the distance from the element $\varphi \in H$ to the sub-space $W_l$.
\end{lemma}
In other words, the space spanned by the first $l$ eigenfunctions of the POD operator provides the best approximation to $f$ in parametric mean distance, among all subspaces of $H$ of dimension $l$ (or smaller).

 From a practical standpoint, to construct the POD of the vademecum (for instance for infected people) it turns out that the correlation matrix $C \in \mathbb{R}^{N \times N}$ relative to the euclidean product in the parameter space,
 $$
 C_{i j} =\Delta t\, \sum_{p=1}^P I_p^i \, I_p^j,\quad i,j=1,\cdots,K
 $$
 is the representation matrix of the POD operator $A$ with respect to the canonical basis $\mbox{   of } H=\mathbb{R}^{K}$.  Therefore, if we consider the diagonalization of matrix $C$,
  $$
 C=V^t D V,
 $$
 --where $D \in \mathbb{R}^{N \times N}$ is a diagonal matrix and $V \in \mathbb{R}^{N \times N}$ is an orthogonal matrix--, then the diagonal elements of D are the eigenvalues of the POD operator $A$ (that we assume to be ordered in decreasing value), and the columns of $V$ are the associated eigenvectors. 
 
 Then we extract as reduced basis the set $\{v^1,\cdots,v^N\}$ with $N$ such that $\lambda_N < \varepsilon$ for a preset threshold $\varepsilon$. Typically we take $\varepsilon = 10^{-6}$. This amounts to $N$ ranging from 2 to 13, the reduced basis really has a very small dimension if we consider that typically the vademecum contains several thousand functions. 
 
\section{Data assimilation by least squares fitting}\label{se:lsq}
The last step of the procedure to obtain a prediction of the evolution of the epidemic is to approximate the real epidemic data by least squares fitting on the reduced space
$$
S_N=\mbox{Span}\{v^1,\cdots,v^M\}.
$$
If the data are known in the discrete times $t_1,\cdots, t_k$ (with $t_k < T$), we use the discrete $L^2(0,t_k)$ norm to assimilate the data:
$$
\|v\|^2_a=\Delta t \, \sum_{m=0}^k |v_m|^2,\quad \forall   v \in \mathbb{R}^M.
$$

We thus search for $I_N \in S_N$ such that
$$
\|I_D-I_N\|_a \le \| I_D- J\|_a,\,\,\, \forall J\in S_N,
$$
where $I_D \in \mathbb{R}^{k+1}$ are the data for infected people at times $t_0,\cdots, t_k$. The same procedure is used to assimilate the data $R_D$ of recovered people.

As the solution $I_N$ is defined for all times $t_k$ for $k=0,\cdots, N$, the values $I_N^i$ for $i>k$ are used to estimate the values of infected people at later times. Similarly the values $R_N^i$ for $i>k$ are used to estimate the future values of recovered people. 

The length of the time period during which these estimates will be accurate will depend on the time smoothness of the function to fit (either $I$ or $R$), on the accuracy of the approximation of the parametric solutions of the SIR model provided by our discretization of the parameter set, on wether our parameter set contains the parameters governing the epidemic evolution, and finally on how accurately the SIR model approximates the evolution of the pandemic.

\section{Validation} \label{se:validation}
\subsection*{Test 1: Validation with analytic functions}
We have initially considered some analytic functions to validate the data assimilation procedure, and determine the range of time in which it may provide a good approximation, beyond the time interval used to assimilate the data. These are a trigonometric and a Gaussian function,
\begin{eqnarray}
&&f_T(t)= \sin(t) + \cos(t) \label{funtrig}\\
&& f_G(t)= \exp \left ({-\left(\frac{t-t_0-1.5}{10}\right )^2} \right )
\label{funexp}
\end{eqnarray}
We have respectively used the following vademecums to fit them:
\begin{eqnarray}
&&F_{2k-1}(t)= \sin(k\,t), \quad F_{2k}(t)=\cos(k\,t)\,\quad\mbox{for  }  k=1,\cdots, m;\label{funtrigv}\\
&& F_{l,k}(t)= \exp \left ({-\left(\frac{t-t_0-l}{k}\right )^2} \right )\,\,\mbox{for  }  l=-m,\cdots, m,\,\, k=5,\cdots, m+5,
\label{funexpv}
\end{eqnarray}
for a given integer $m\ge 1$. 

We present in Figures 1 and 2 the results for the first case. We intend to approximate function $f_T$ in $[0,2\,\pi]$.  As $f_T$ belongs to the vademecum, it should be exactly approximated when we provide data enough. We set $m=5$, then the vademecum contains $P=10$ functions, and provide the values of these functions at equally spaced discrete times $t_k$ with step $2\,\pi/100$. We provide data to assimilate for $k=1,\cdots, 16$. We observe that for $N=10$ eigenvectors the fitting is quite accurate in the whole interval $[0,2\,\pi]$, while for $N=9$ there is a large error. Providing data for larger times ($k\ge 17$) for $N=10$ yields a perfect fitting in the interval $[0,2\,\pi]$, but it only slightly improves the fitting for $N\le 9$. However if the data grid is refined, this behavior improves: Figure 2 displays the fitting with $9$ eigenfunctions with a grid of half step, using data in the interval $[0, 6\,\pi/5]$. 

In all cases the fitting in the time interval were data are provided is quite accurate.
%\vspace*{-2,8cm}
\begin{figure}[htb]
\centering
\subfigure{\includegraphics[width=80mm]{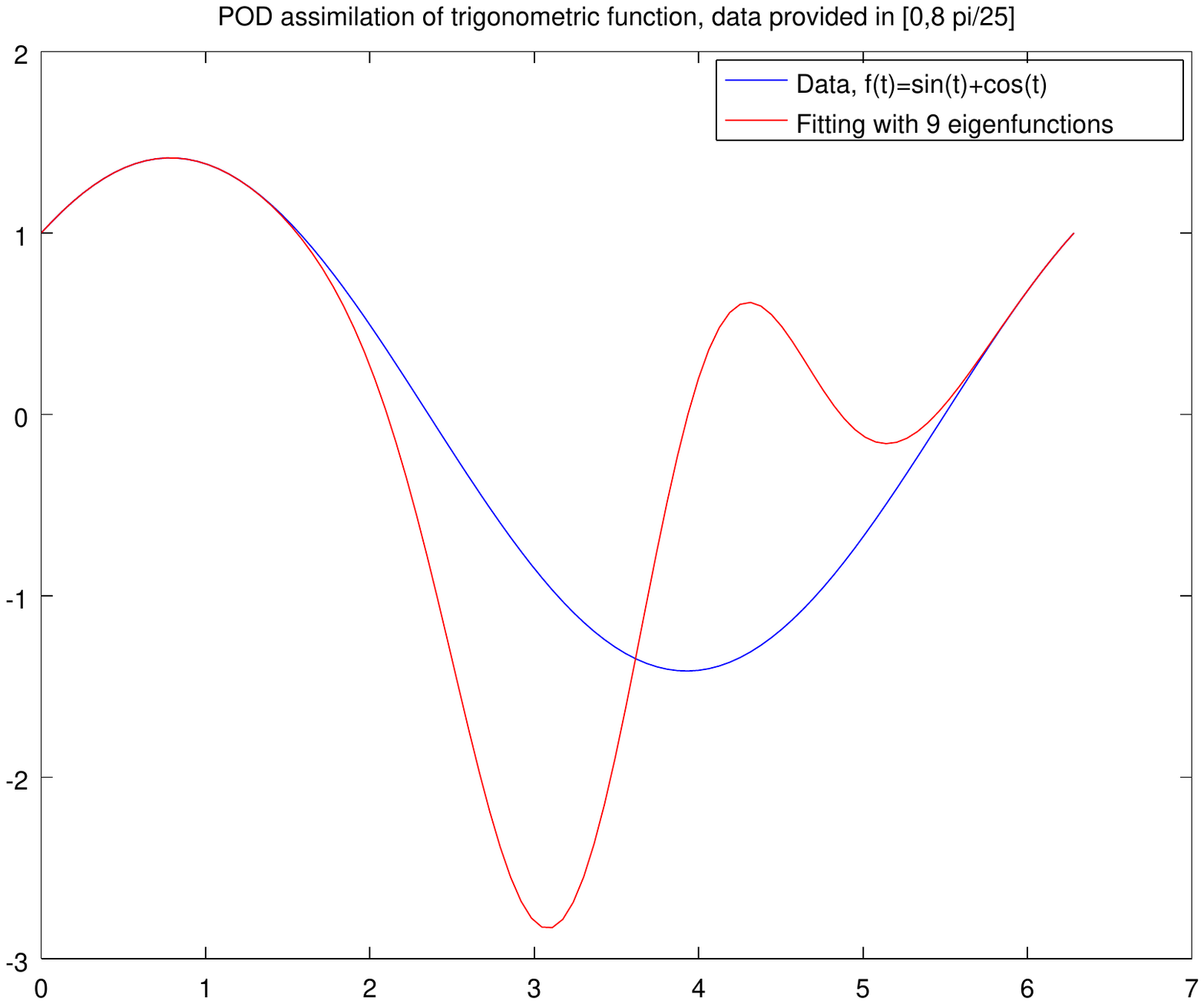}}
\subfigure{\includegraphics[width=80mm]{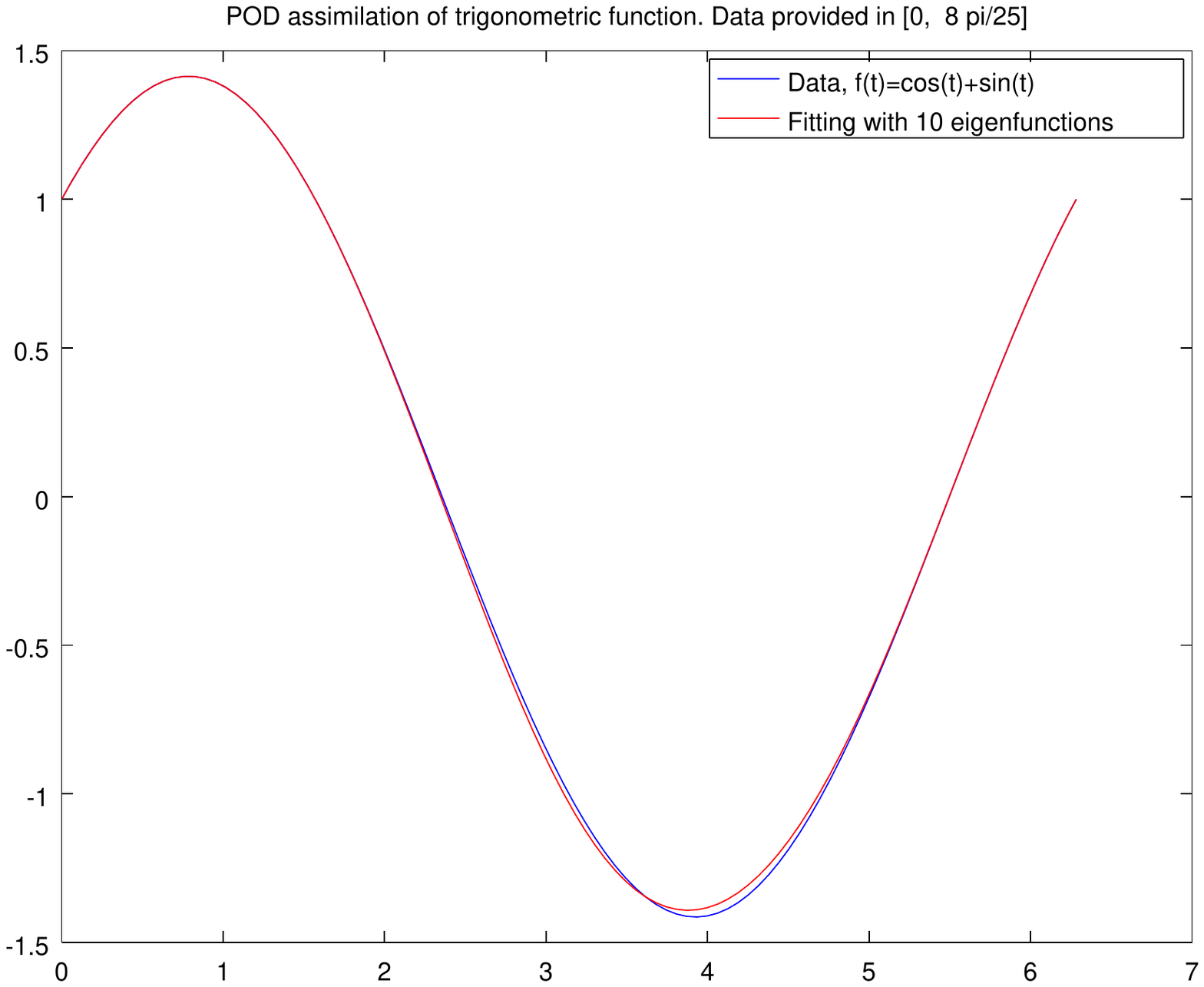}}
\vspace*{-2.8cm}
\caption{Test 1: Data assimilation for trigonometric function} \label{fig:asimtrig}
\end{figure}

%\vspace*{-2,8cm}
\begin{figure}[htb]
\centering
\subfigure{\includegraphics[width=80mm]{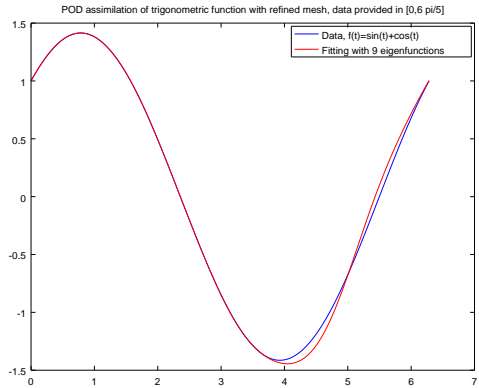}}
\vspace*{-2.8cm}
\caption{Test 1: Data assimilation for trigonometric function. Refined mesh data.} \label{fig:asimtrig2}
\end{figure}

Figures 3 and 4 display the results for the second analytic test  \eqref{funexp}-\eqref{funexpv}. We intend to approximate the function $f_G$ in the interval $[0,120]$. We set $t_0=55$ and $m=3$, what yields $P=28$ vademecum functions. We provide the values of the vademecum functions at equally spaced discrete times $t_k$ with step $120/100$. We provide $t_k$ for relatively small intervals $[0,57]$ and $[0,60]$. We observe that the fitting improves as the interval where data are provided increases, and a large gain in accuracy can be obtained with few more data (Figure 3). However if this interval is not large enough (Figure 4) the fitting does not improve as the number of eigenmodes increases. For a relatively reduced interval where data are provided, the best fit is provided by a moderate number of eigenmodes. Furthermore, large oscillations of the fitting appear when the number of eigenmodes is excessive. This possibly occurs because the large frequencies are not well fitted. 

%\vspace*{-0.8cm}
\begin{figure}[htb]
\centering
\subfigure{\includegraphics[width=80mm]{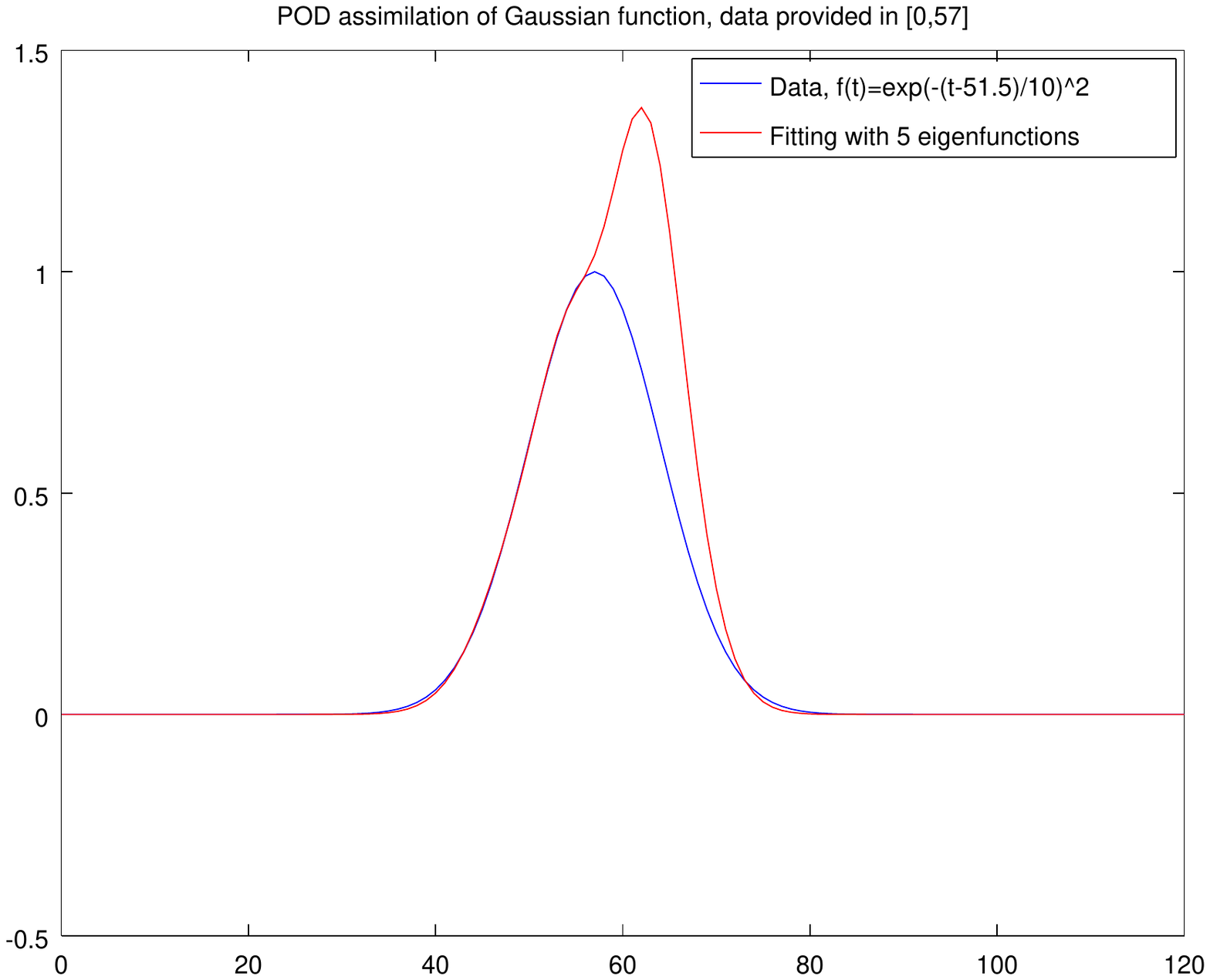}}
\subfigure{\includegraphics[width=80mm]{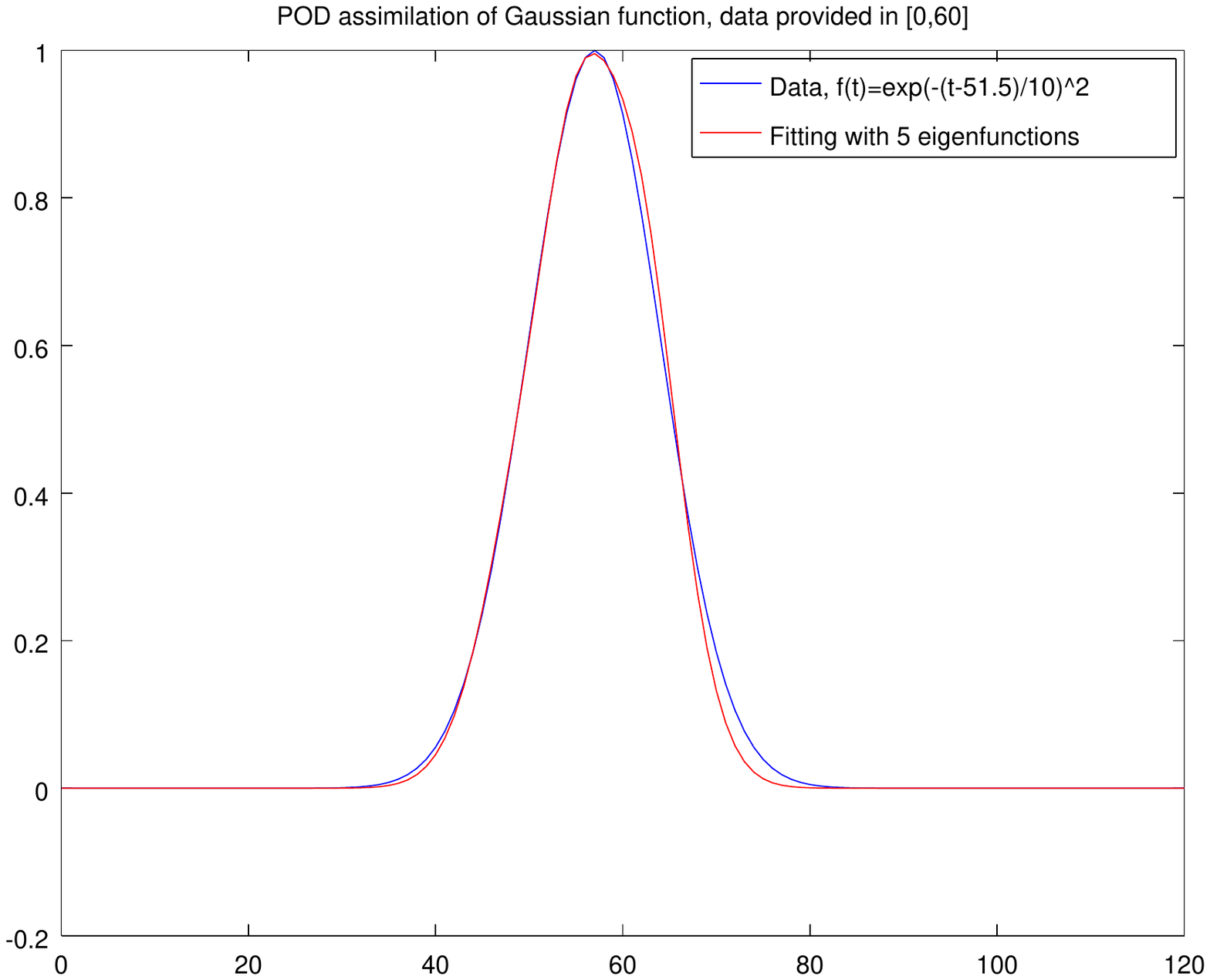}}
\vspace*{-2.8cm}
\caption{Test 1: Data assimilation for Gaussian function. Fitting with 5 eigenfunctions.} \label{fig:asimtrig3}
\end{figure}

In both tests we have observed that large oscillations of the fitting function, beyond the interval in which data are given, appear when the number of assimilated data is not large enough for a given amount of eigenfunctions. 

\subsection*{Test 2: SIR model solution}
This test is intended to analyze the ability of the ROM assimilation procedure to accurately predict the solution of the SIR model. 

%\vspace*{-1,8cm}
\begin{figure}[htb]
\centering
\subfigure{\includegraphics[width=80mm]{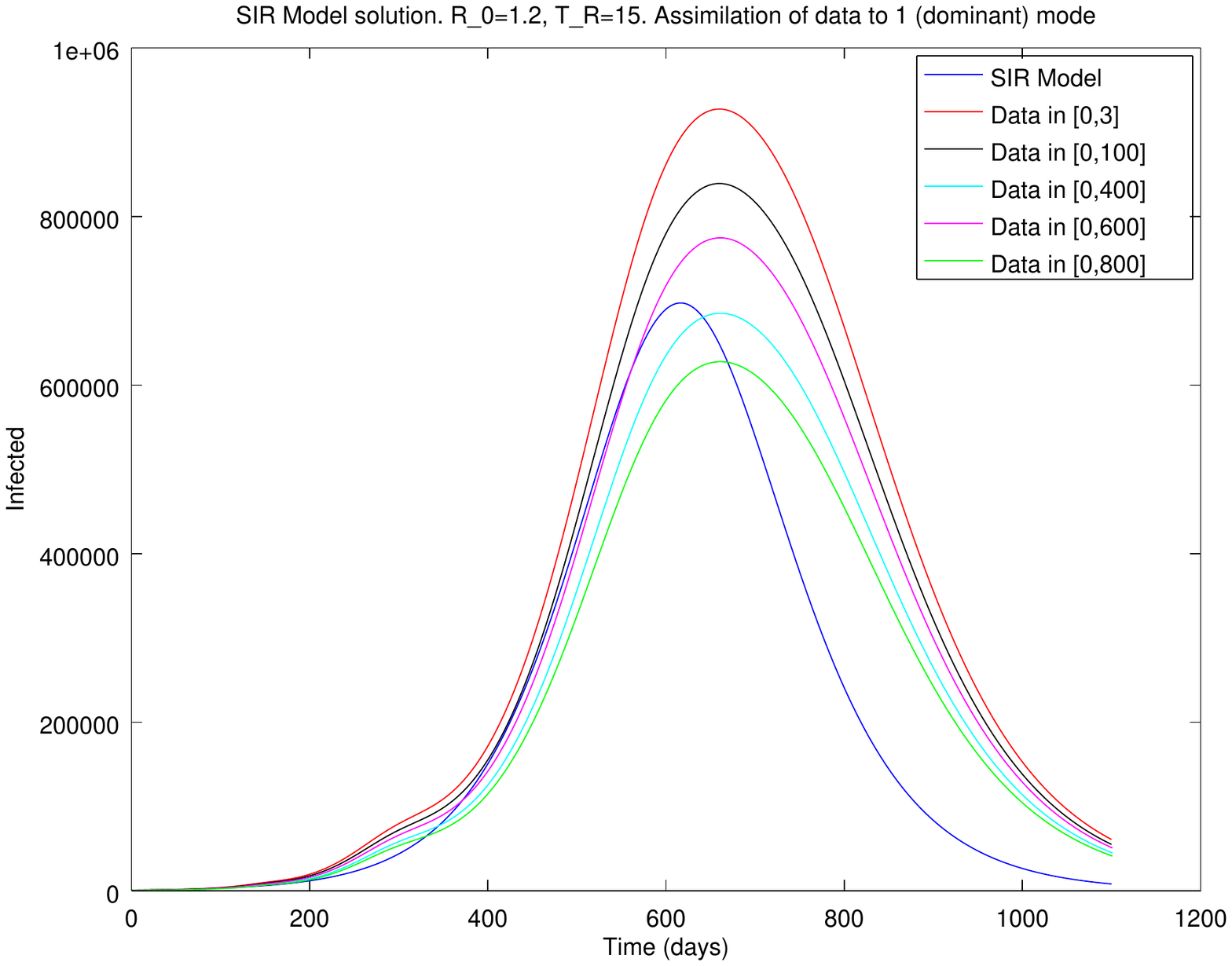}}
\subfigure{\includegraphics[width=80mm]{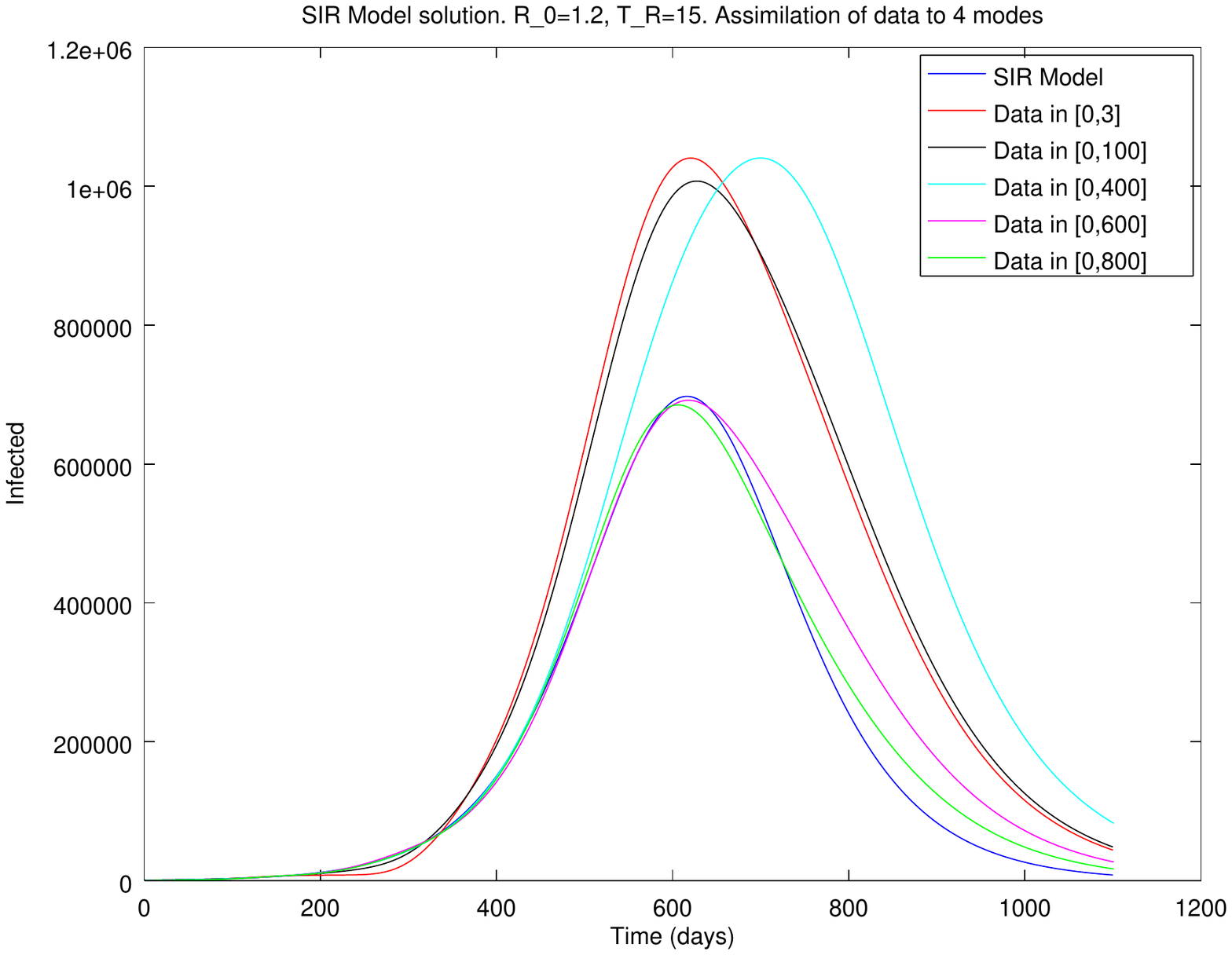}}
\vspace*{-2.8cm}
\caption{Test 2: Data assimilation for SIR model solution to $N=1$ (left) and $N=4$ (right) POD modes.} \label{fig:asimdom}
\end{figure}

With this purpose we have constructed the vademecum by solving model \eqref{modsir2} in the time interval $[0,1100]$ and time step $1$ day, with data $S_\infty= 47.100.396$ (official Spanish population in 2020), and $I_0$, $R_0$,  $r_0$ and $T_r$ respectively ranging in the sets $\{800,\, 900,\, 1000\}$, $\{50,\, 60,\, 70\}$, $\{0.6,\, 0.8,\, 1.1,\, 1.3\}$, and $\{5,\, 10,\, 20,\, 25\}$. We have tested the length of the time interval in which data must be feed to the assimilation algorithm, as well as the number of modes needed to obtain accurate results.

%\vspace*{-1,8cm}
\begin{figure}[htb]
\centering
\subfigure{\includegraphics[width=80mm]{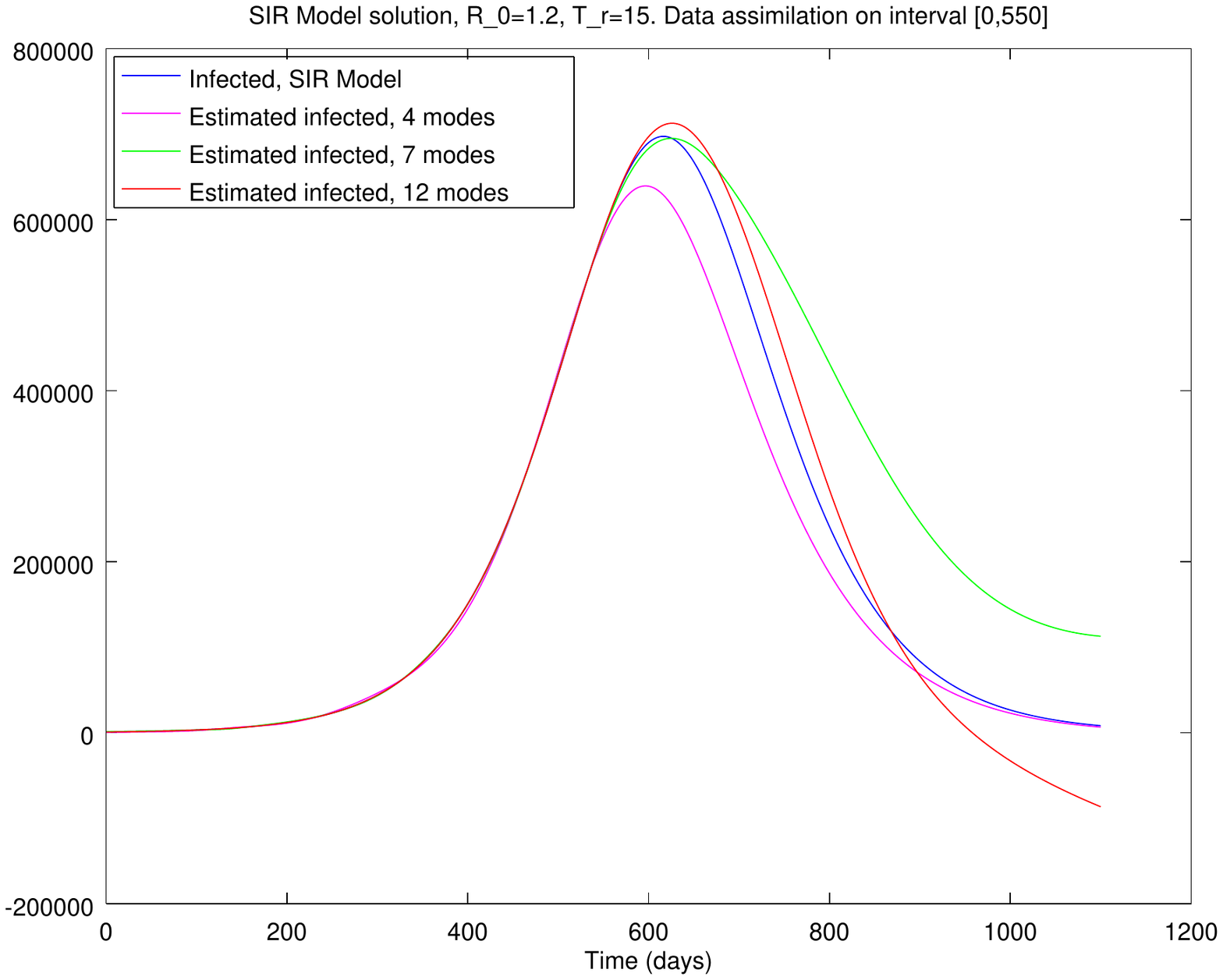}}
\subfigure{\includegraphics[width=80mm]{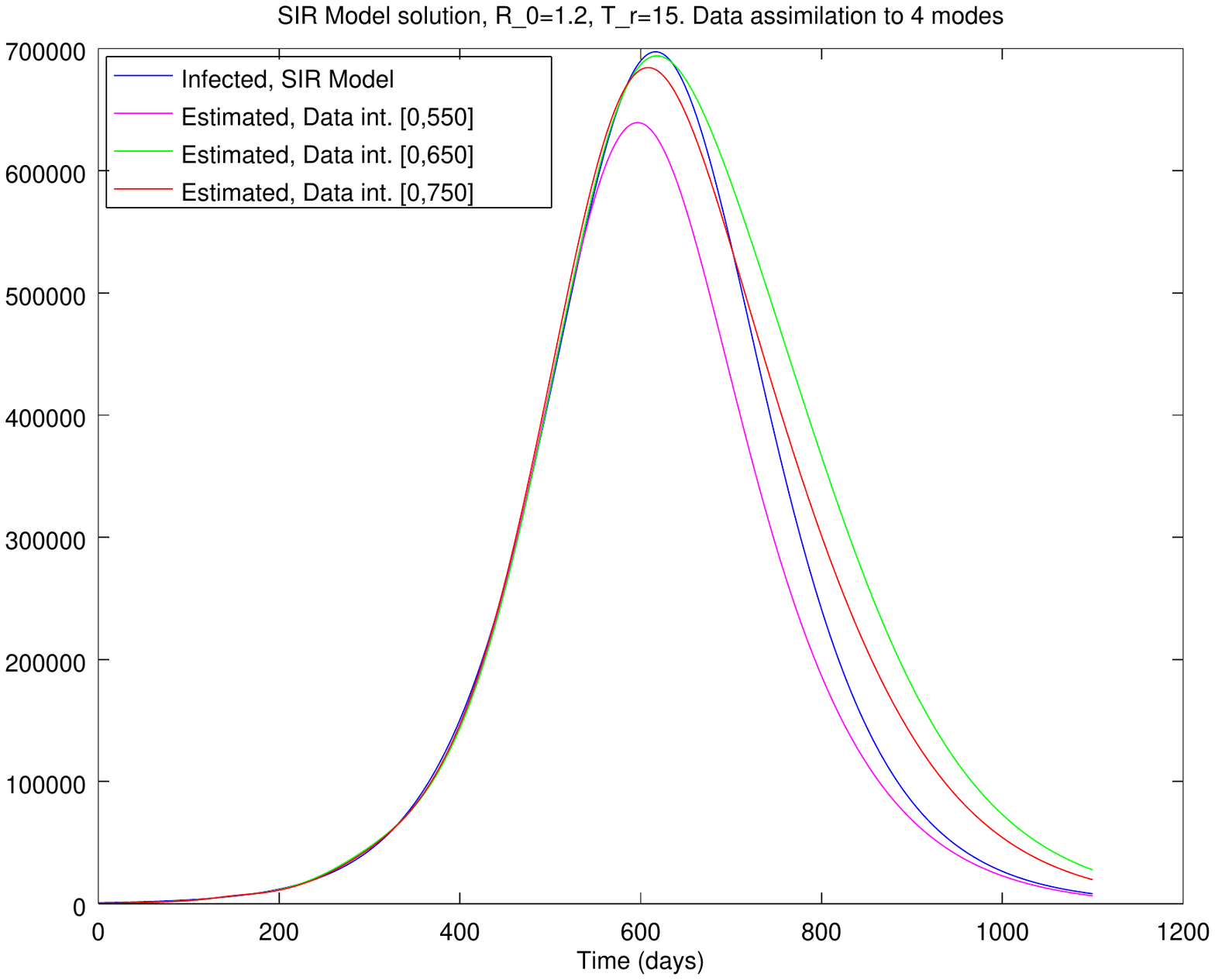}}
\vspace*{-2.8cm}
\caption{Test 2: Data assimilation for SIR model solution to several number of POD modes on fixed data assimilation interval (left), and to a fixed number of modes on variable data assimilation intervals (right).} \label{fig:asimvariosn}
\end{figure}

Figure \ref{fig:asimdom} shows the assimilation to $N=1$ (dominant mode) and $N=4$ modes, of the data for infected people with different intervals of data assimilation. The qualitative behavior of the pandemic is roughly well approximated when $N=1$, with estimated number of maximum number of infected people between 70\% and 130\% of those given by the SIR model. In addition, the estimation of the time at which the number of infected people reach its maximum is well approximated when $N=4$, with increasing global accuracy as the time data interval increases. Assimilating data with as few data as $t_1$ to $t_4$ (and not less) allows to reproduce the overall features of the pandemic. This provides data to be assimilated for 4 discrete times, right the number of parameters on which depends problem \eqref{modsir2}. However gaining in accuracy for this vademecum requires a large amount of data, at least for $[0,550]$, which is almost half of the time interval in which the epidemic experiences appreciable changes (see Table \ref{tabla:sencilla}).

\begin{table}[htbp]
\begin{center}
\begin{tabular}{|c|c|c|}
\hline
Data interval & Numer of modes & Relative error \\
\hline \hline
[0, 3]&	4&	65'55\% \\ \hline
&	11&	99'92\% \\ \hline
&	12&	Singular Gram Matrix \\ \hline\hline
[0, 550]&	4&	13'2\% \\ \hline
&	11&	10'7\% \\ \hline
&	12&	11'4\% \\ \hline\hline
[0, 750]&	4&	8'7\% \\ \hline
&	11&	16'9\% \\ \hline
&	12&	27'8\% \\ \hline\hline
[0, 1.100]&	4&	5'4\% \\ \hline
&	11&	2'3\% \\ \hline
&	12&	1'9\% \\ \hline\hline
\end{tabular}
\caption{Test 2: Errors for data assimilation for SIR model solution to several number of POD modes and increasing time data intervals. The relative error is calculated in discrete $l^2(0,1.100)$ norm}
\label{tabla:sencilla}
\end{center}
\end{table}

Figure \ref{fig:asimvariosn} exhibits the comparison of the fitting provided for several number of modes on a single time interval of data, and for $N=4$ modes with different data assimilation intervals. We observe that in all cases the accuracy is improved with respect to the fitting with 1 mode for a number of modes between 1 and 4. However a number of modes above 4 produce less accurate approximations, in some cases with negative values. This is likely due to the high frequency components that are not well fitted unless the data are provided in a large enough interval, similarly to what happened in Test 1. The best result correspond to a relatively reduced number of modes ($N=4$). Increasing the fitting data interval progressively decreases the error, that also is smaller for larger number of modes, decaying to values below 2\%. This may also be observed in Table 1. Note that the error for $N=12$ modes is larger than the one for $N=11$ modes until the time interval is nearly the full computation time interval.

\begin{figure}[htb]
\centering
\subfigure{\includegraphics[width=80mm]{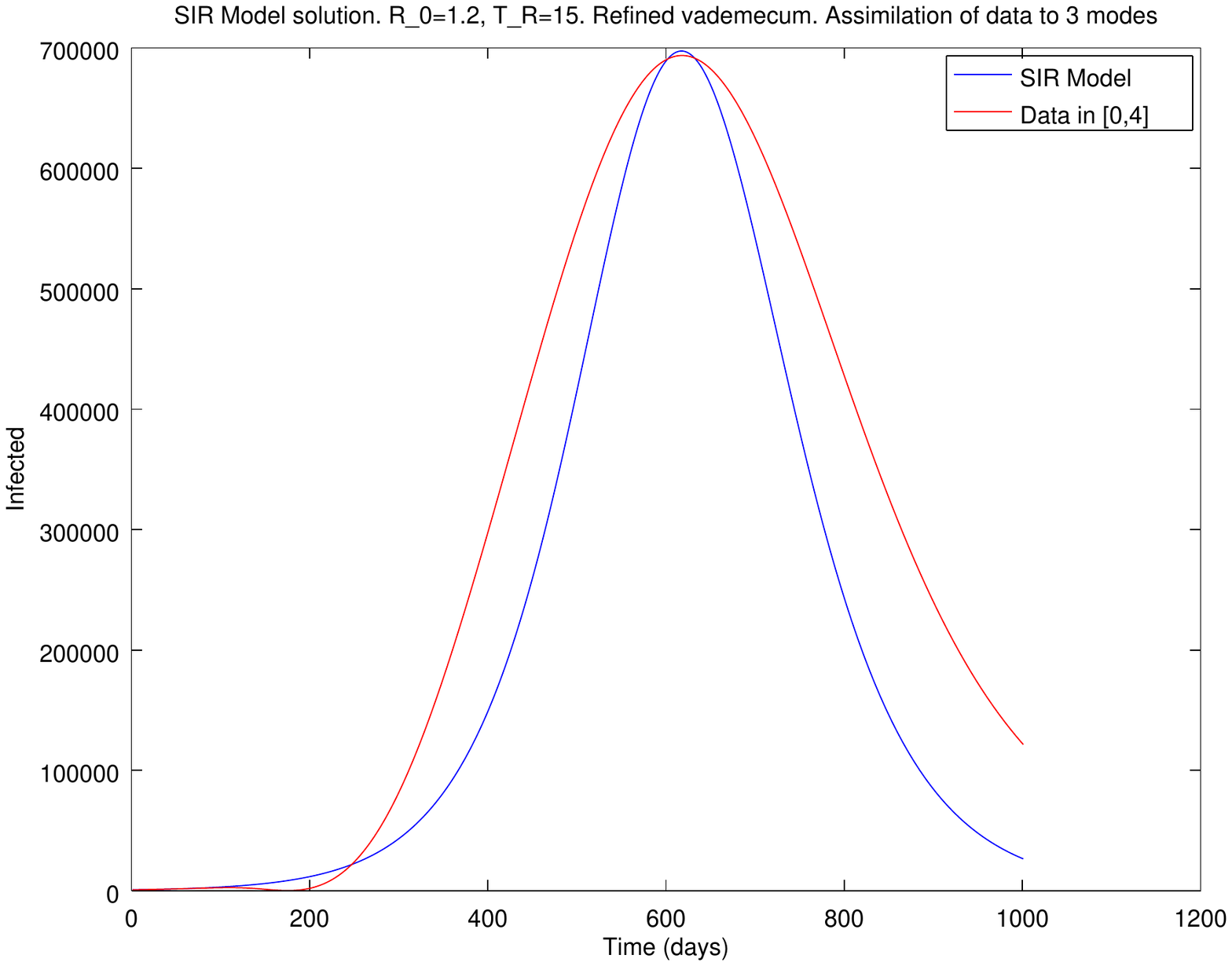}} \subfigure{\includegraphics[width=80mm]{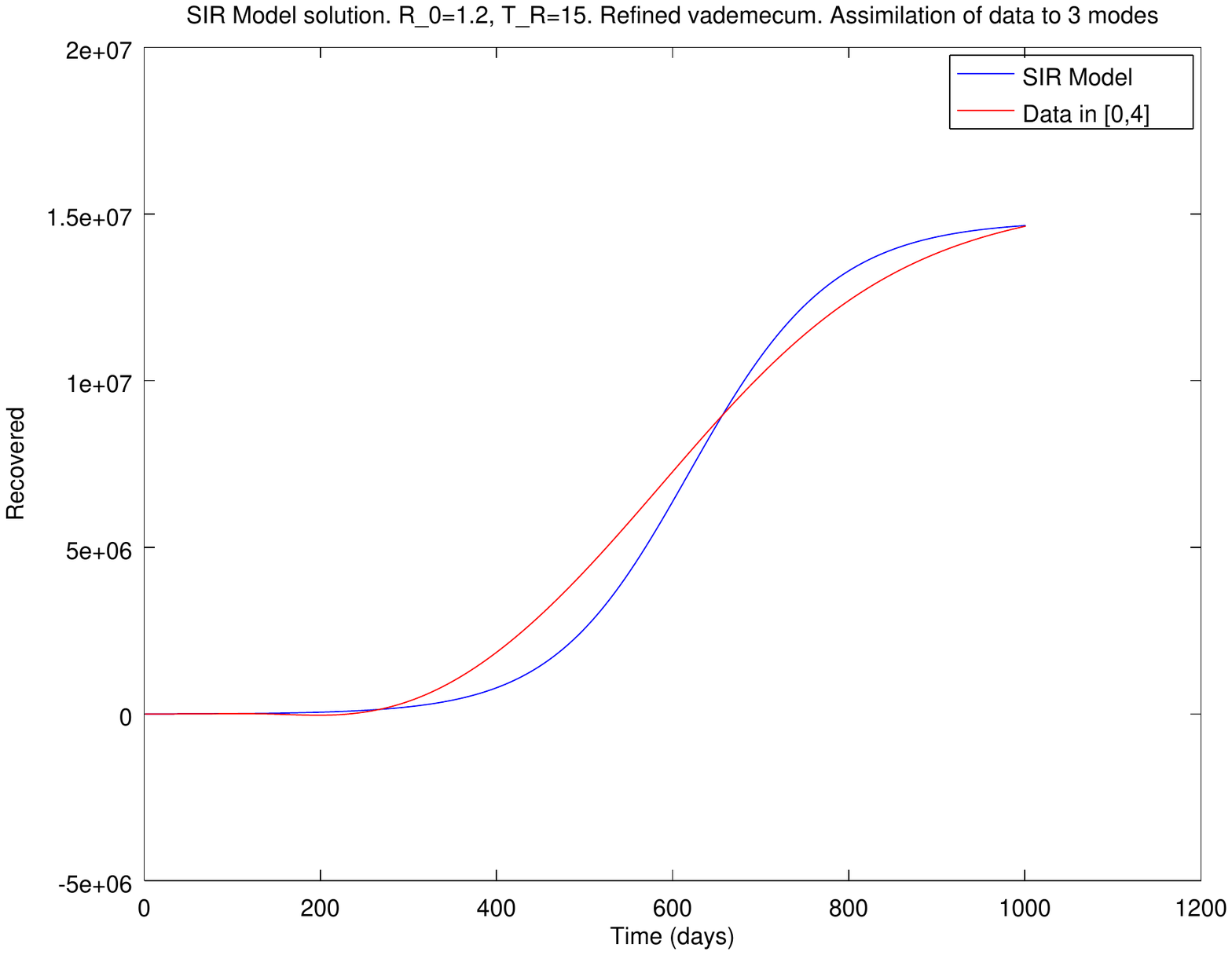}}\\
\vspace*{-2.8cm}
\caption{Test 2: Data assimilation with vademecum constructed with refined data grid, for infected (left) and recovered (right) populations.} \label{fig:refinedfit}
\end{figure}

\begin{figure}[htb]
\centering
\subfigure{\includegraphics[width=80mm]{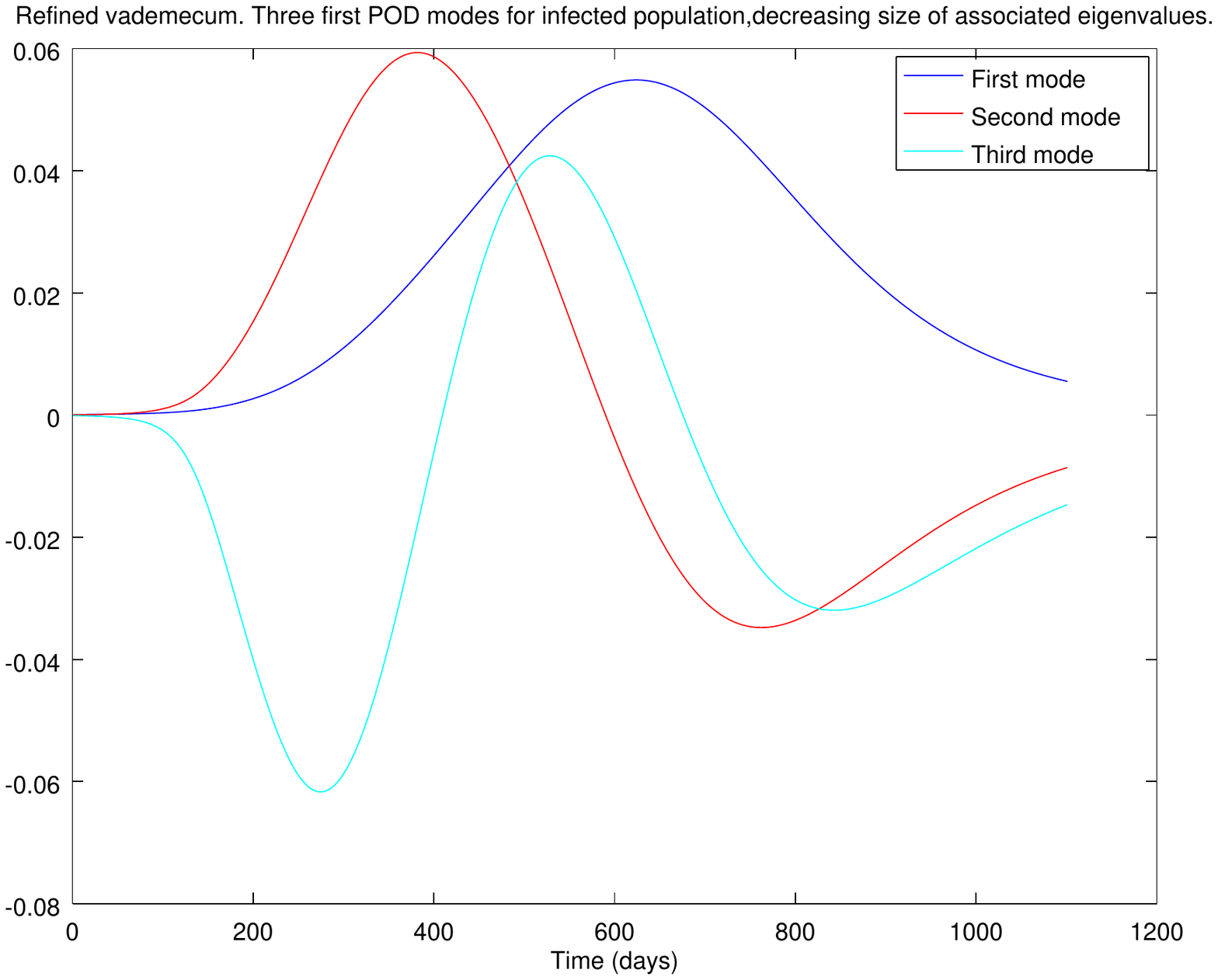}}
\subfigure{\includegraphics[width=80mm]{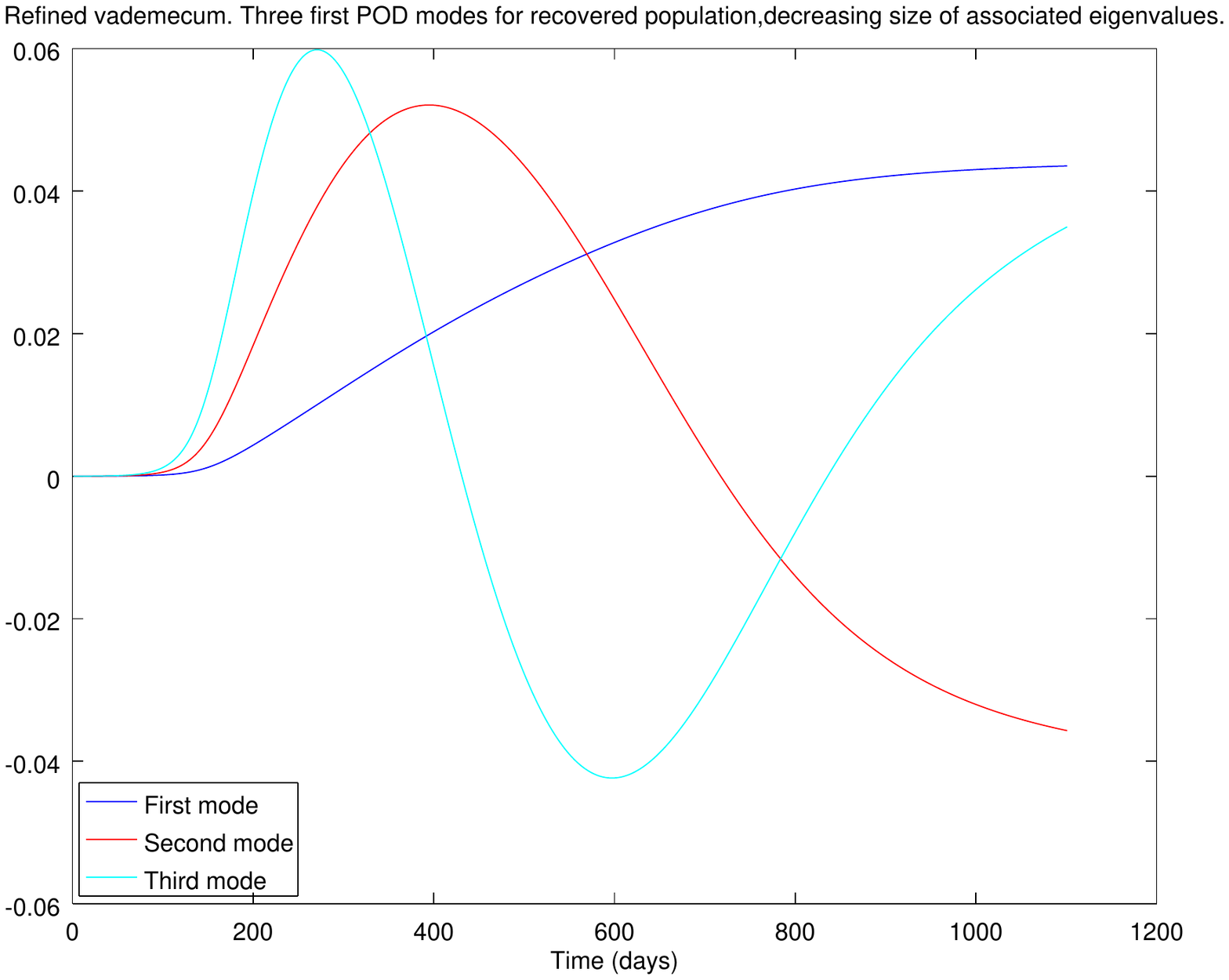}}
\vspace*{-2.8cm}
\caption{Test 2: Data assimilation with vademecum constructed with refined data grid. First three dominant eigenvalues for infected (left) and recovered (right) populations.} \label{fig:refinedfit2}
\end{figure}

An alternative, and feasible, way to increase the accuracy is to enrich the vademecum with additional solutions of the SIR model, for finer grids of data. We have assimilated the SIR solution data obtained for $I_0=850$, $R_0=55$, $r_0=1.2$ and $T_r=15$, as before, but now on the reduced POD space constructed with the vademecum corresponding to the data $I_0=850$, $R_0=55$ and  $r_0$ and $T_r$ respectively ranging in the sets  $\{0.6, 0.65,\cdots,1.15 \}\cup \{1.25, 1.30\}$, and $\{5,5.5,\cdots,14.5\}\cup\{15.5,16,\cdots,24.5,25\}$. Figure \ref{fig:refinedfit} displays the assimilation of just 4 data values, at times $t_1$ to $t_4$. A good accuracy for the overall evolution of infected and recovered population is obtained, if the data are assimilated to as low as $N=3$ POD eigenmodes. Figure \ref{fig:refinedfit2} shows the 3 dominant eigenmodes, note that the shape of the first eigenmode retains the dominant pattern of the pandemic evolution.

In all tests performed up to now we have observed that large oscillations of the fitting function, beyond the interval in which data are given, appear when the number of data is not large enough for a given amount of eigenfunctions. This anomalous behavior will be used as an indicator to avoid un-accurate fitting. 
\begin{figure}[htb]
\centering
\subfigure{\includegraphics[width=52mm]{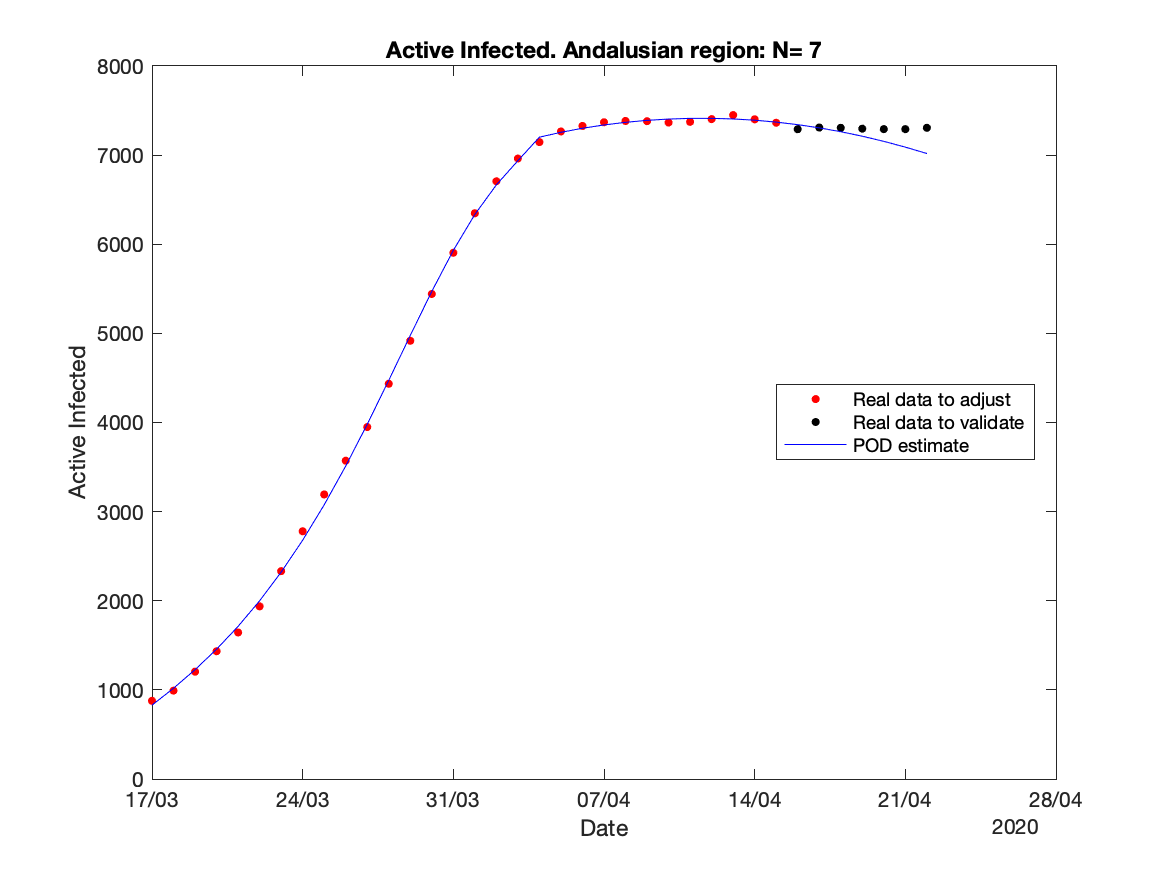}}
\subfigure{\includegraphics[width=52mm]{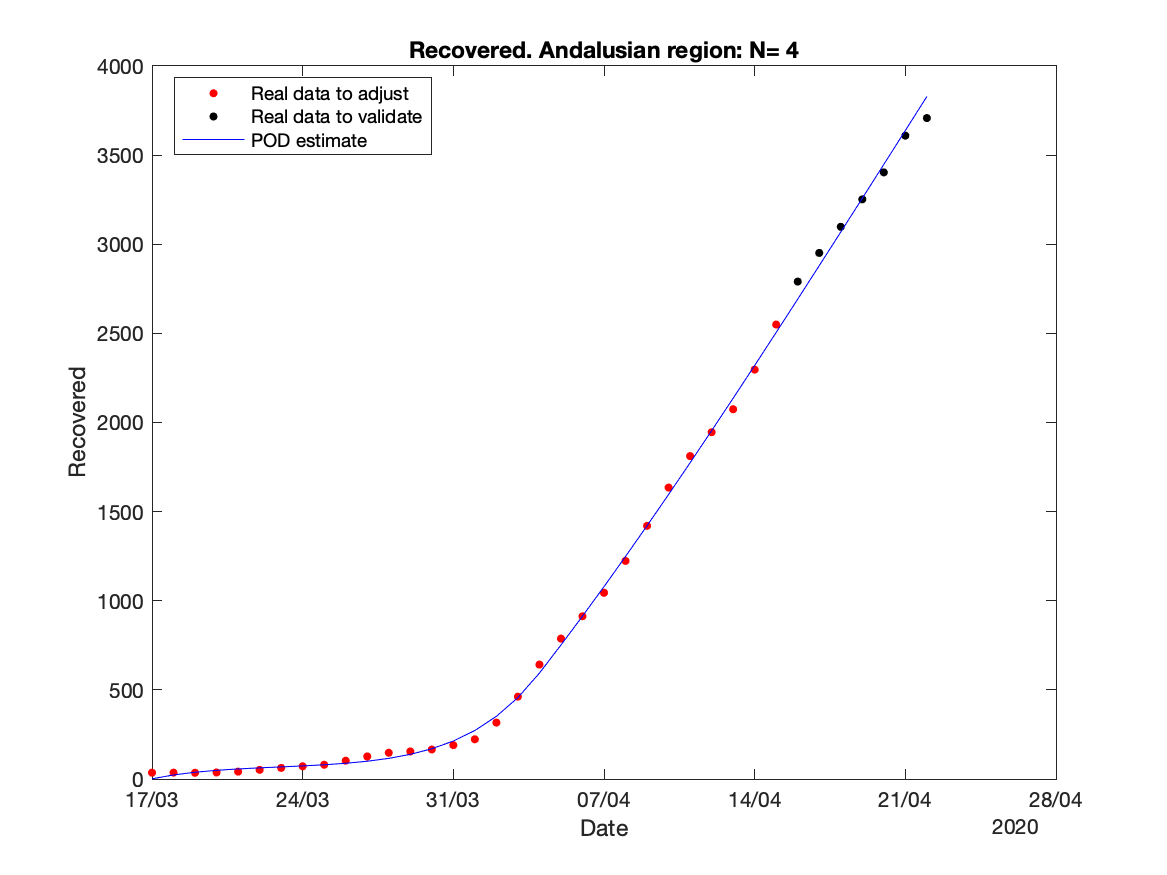}}\subfigure{\includegraphics[width=52mm]{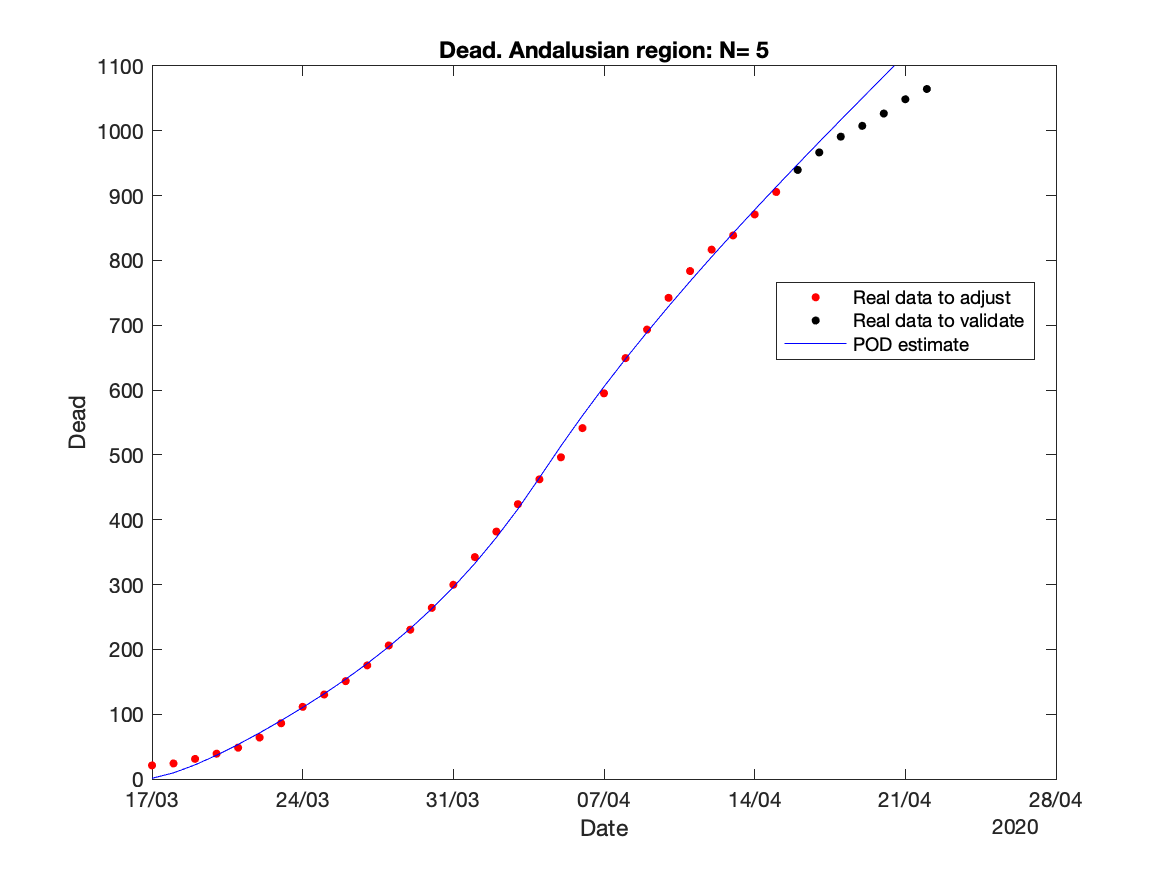}} \\\subfigure{\includegraphics[width=52mm]{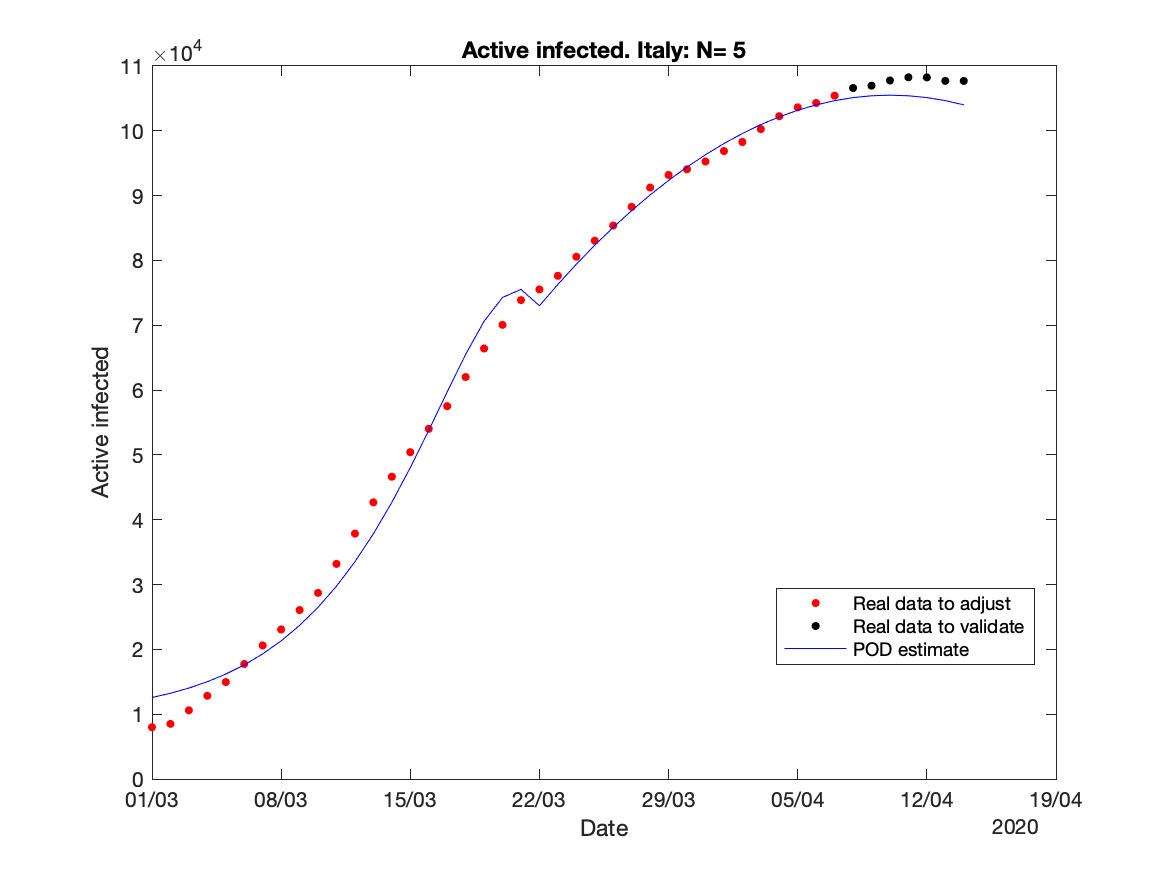}}
\subfigure{\includegraphics[width=52mm]{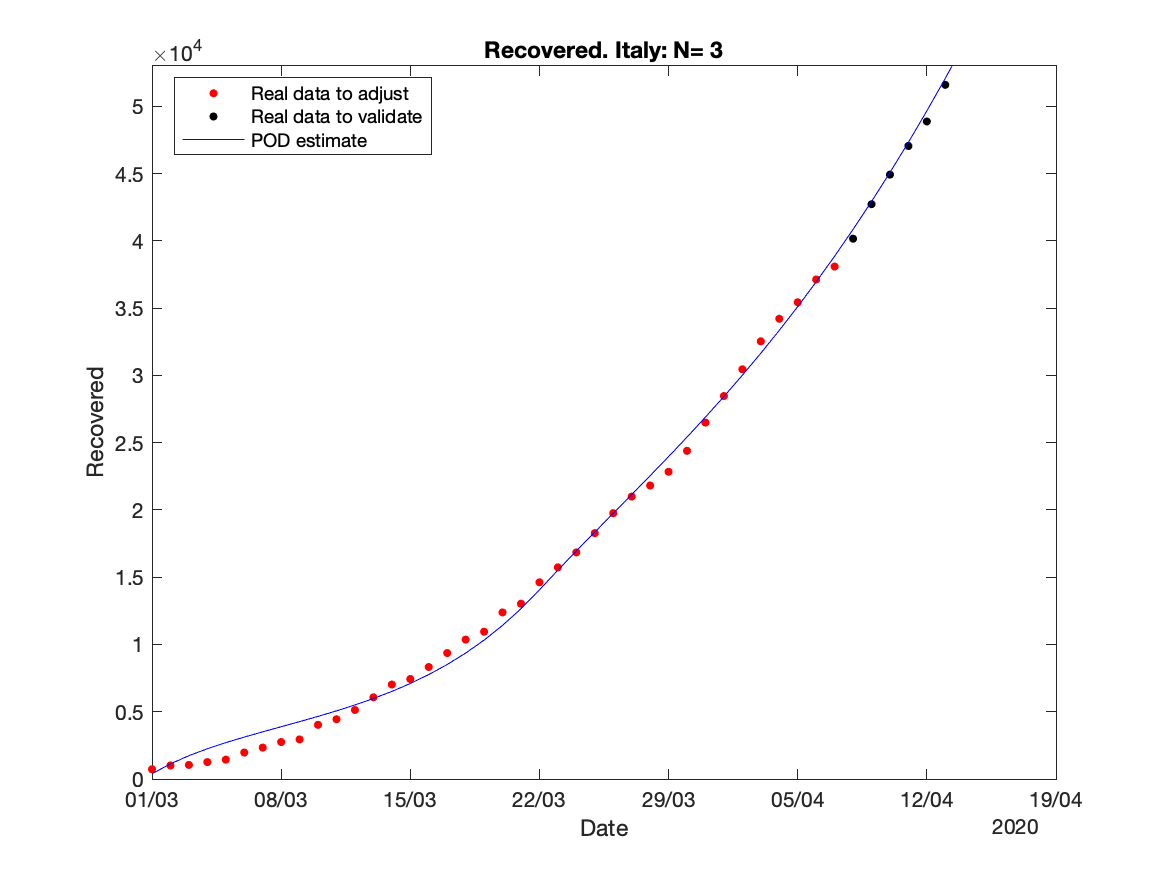}}\subfigure{\includegraphics[width=52mm]{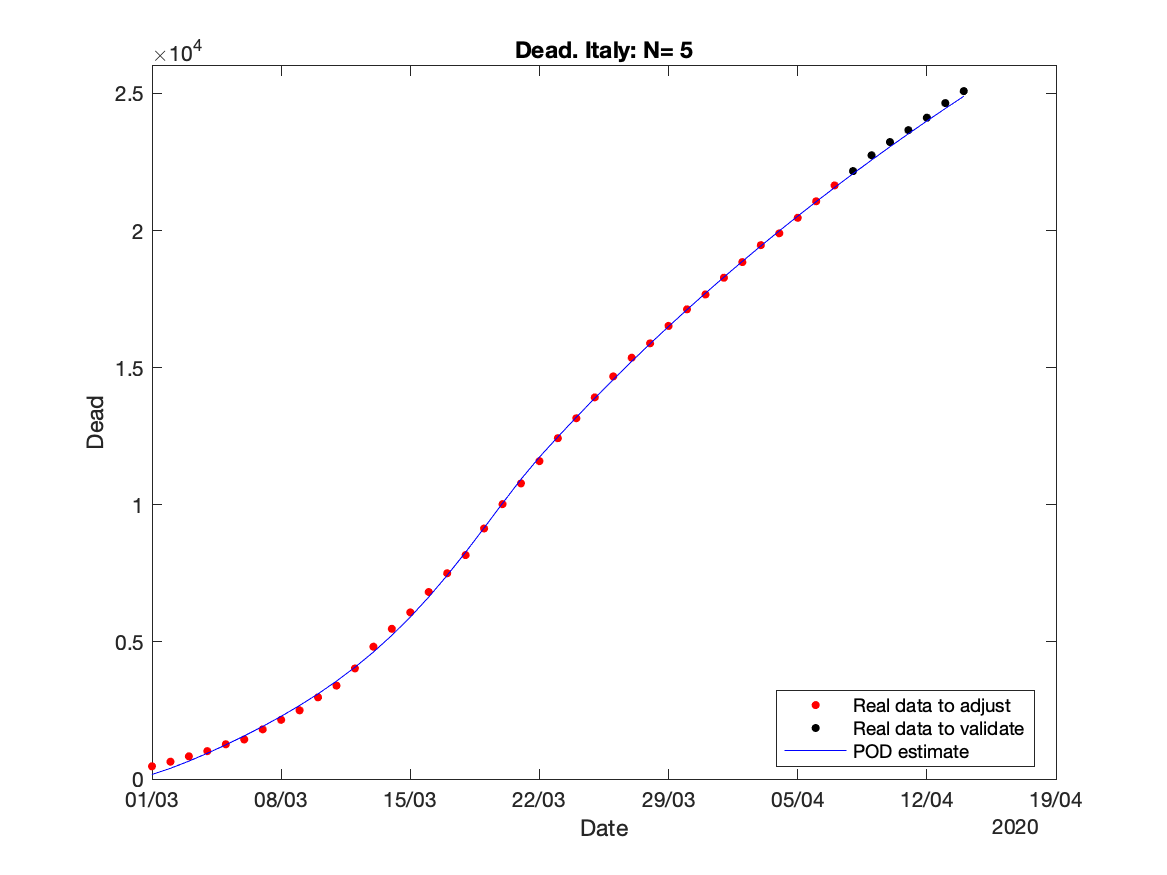}}\\\subfigure{\includegraphics[width=52mm]{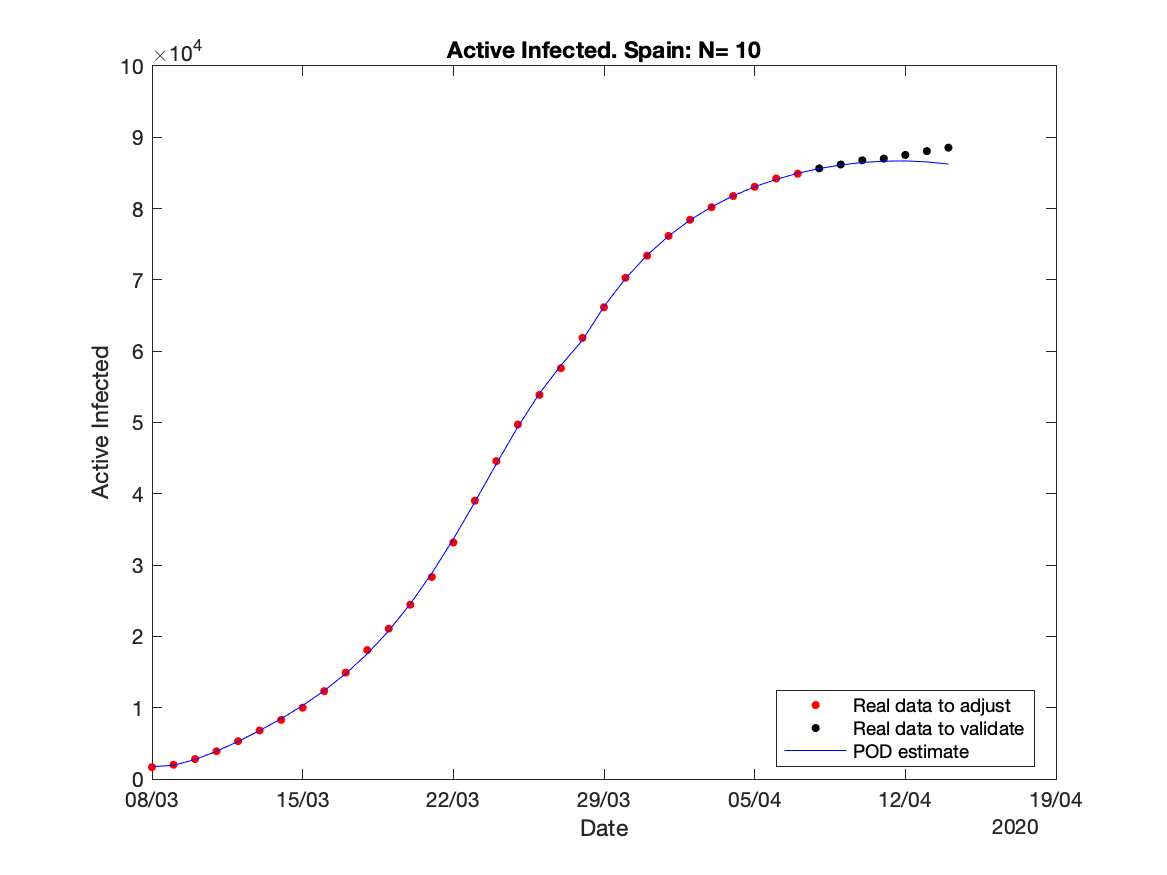}}
\subfigure{\includegraphics[width=52mm]{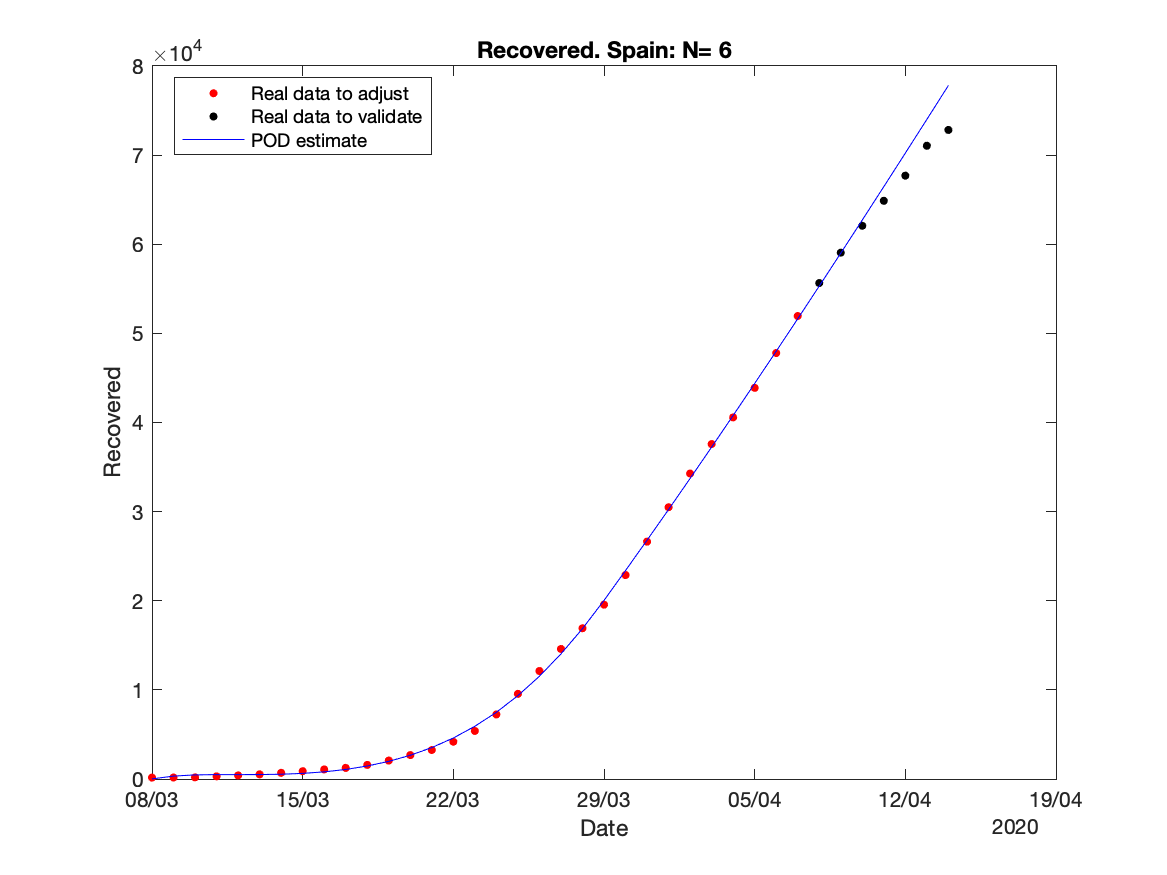}}\subfigure{\includegraphics[width=52mm]{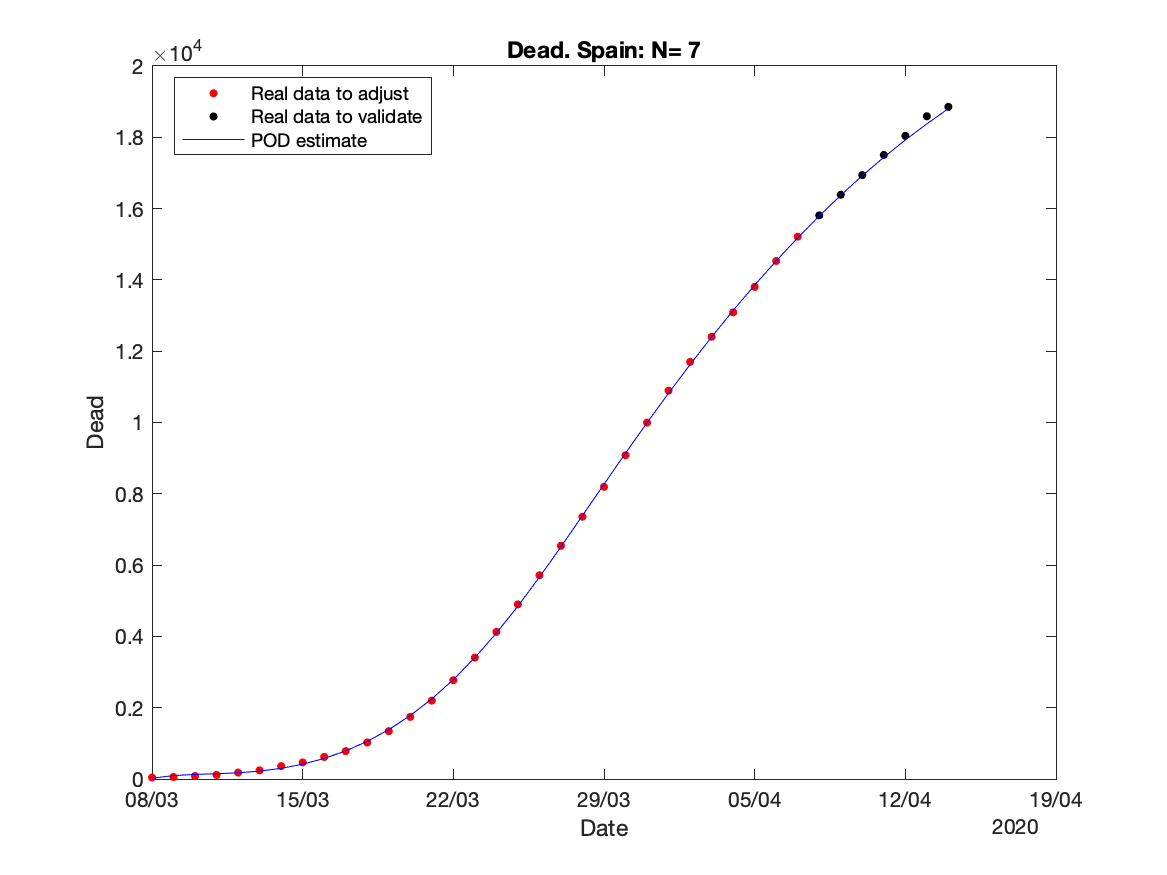}}
\caption{Test 3: Prediction of Covid-19 pandemic evolution in Andalusia region, Italy and Spain to 7 days. Red dots represent the assimilated data, blue dots the predicted ones, and the blue lines are the fitted curves. The peaks in these curves correspond to the change in the basic reproduction rate $r_0$.} \label{fig:predandalucia}
\end{figure}

\subsection*{Test 3: Covid-19 pandemic}
We have applied the ROM data assimilation procedure to the Covid-19 pandemic with data from Italy, Spain and the Andalusia region in Spain. 

As starting time we take the day when the lockdown took place in each area (March 9 2020 in Italy and March 16 in Andalusia and Spain). To construct the vademecum, as we mentioned in Section 2 we typically take 4 values for the initial conditions for the parameters, equally spaced, neighboring the reference values. We also assume the basic reproductive rate $r_0$ to take two values in different time intervals, as indicated in \eqref{r0variable}, to take into account the relaxation in meeting the lockdown measures. The reference times $T_1$ corresponds to the days in which the lockdown measures appeared to relax in each country or region: $T_1=$ April 4 for Andalusia Region and March 30 for Italy and Spain. The initial data have been taken at March 17 for Andalusia and March 9 for Italy and Spain. The computation time $T$ is twice the time in which assimilation data are provided.

\begin{table}[htbp]
\begin{center}
\begin{tabular}{||c|c||c|c||c|c||}
\hline
 \multicolumn{2}{|c||}{}&\multicolumn{2}{|c||}{\bf Prediction to 5 days}&\multicolumn{2}{|c||}{\bf Prediction to 7 days}\\\hline
\hline 
Area &Population & Number of modes&  Relat. error & Number of modes&  Relat. error \\
\hline \hline
\bf Andalusia	&Infected & 7 &4'60\%\%& 7 &3'94\%	 \\ \hline
                      &Recovered& 4 &	6'95\%& 4 &	3'25\% \\ \hline
	                &Deceased  &4&	4'43\%&5&	8'09\% \\ \hline\hline
\bf Italy        	&Infected & 5 &2'08\%& 5 &3'41\%	 \\ \hline
                     &Recovered& 3 &	1'47\%& 3 &	1'44\% \\ \hline
	                &Deceased  &5&	0'55\% &5&	0'79\% \\ \hline\hline
\bf Spain		&Infected & 10 &2'61\%& 10 &2'59\%	 \\ \hline
                      &Recovered& 3 &1'29\%& 6 &6'85\% \\ \hline
	                &Deceased  &7&	0'75\%&7&	1'11\% \\ \hline\hline
\end{tabular}
\caption{Test 3: Maximum relative errors for 5 and 7-days prediction of Covid-19 pandemic evolution in Andalusia region, Italy and Spain. }
\label{tabla:errorescovid19}
\end{center}
\end{table}

 To validate the data we use a time period to assimilate the data, and compare the predicted values in a later time period with the official ones. 
 
We have adapted the ROM data assimilation procedure to the prediction of the number of deceased people. We split the recovered $R$ into true recovered $P$ and deceased people $D$, and rewrite the SIR model as 
  \begin{equation} \label{modsird}
 \left \{ \begin{array}{rcl}
 S'&=&-\alpha(t)\, \displaystyle\frac{S}{S_\infty}\, I , \\
 I'&=& \,\,\,\,\displaystyle\alpha(t)\, \frac{S}{S_\infty}\, I - \beta_r \, I -\beta_d \, I\\
 P'&=&\beta_r \, I,\\
 D'&=&\beta_d\, I,
 \end{array}
 \right .
 \end{equation}
 where $\beta_r=\rho \, \beta$ and $\beta_d=(1-\rho)\,\beta$, where $\rho \in (0,1)$ is a new parameter. Given the mortality of the pandemic, we have set $\rho=0.9$ as reference value for this parameter. To compute the vademecum, the parameter $\rho$ is also let to take 4 values equally spaced around the reference value, within the interval $(0,1)$.

 Figures \ref{fig:predandalucia} displays the results for Andalusia region, Italy and Spain, with official data respectively taken from \cite{Andaludat},\cite{Spaindat} and \cite{Italydat}. As data for Andalusia and Spain present large daily oscillations, we have instead used the moving time average over 3 days. We show the comparison of the fitting curve with the official reported values for the last 5 days. Among all possible number of modes, we select those for which the fitting function presents the smallest oscillations after the fitting data interval, without reaching negative values. We observe that the errors within the fitting interval are very small. 
 
 Table \ref{tabla:errorescovid19} displays  the relative errors for infected, recovered and deceased people in the three countries/regions for predictions to the next 5 and 7 days. Both predictions to 5 and 7 days are rather accurate, with errors typically below 4\%, and in all cases below 9\%. Errors for Italy and Spain are smaller, possibly because the populations are larger. The errors corresponding to predictions of infected and recovered populations to 7 days in Andalusia are smaller than to 5 days. This is not inconsistent as we are fitting the last either 5 or 7 days, and assimilating the data in all the preceding days. Some oscillations in a relatively small amount of data to assimilate possibly originate this behavior.
 
 \section{Conclusions}\label{se:conclusions}
 This paper introduces a data assimilation procedure for Covid-19 based upon Reduced Order Modelling (POD) of the solutions of the SIR epidemic model. It is based upon the hypothesis that although the actual model parameters that govern the epidemic are unknown, reference values are known. Thus the actual evolution of the pandemic may be well approximated by the vademecum generated by solving the SIR model with a number of parameters nearby the reference ones.
 
 This procedure has been tested with analytic functions and the solution of SIR model. We have concluded that:
 \begin{itemize}
 \item The fitting error within the data assimilation interval is very small in all cases.
  \item
 Assimilating data in a small time interval (at least as many days as parameters appear in the SIR model) to a very small number of modes provides a qualitative good approximation of the evolution of the epidemic: A rough approximation of the maximum number of infected persons and of the time at which the maximum takes place.
 \item This assimilation of a very small number of data on a reduced space of very small dimension becomes fairly accurate if the vademecum formed with solutions of the SIR model is enriched by refining the parameters grids.
 \item
 It is preferable to assimilate data to a moderate number of modes, that provides the best approximations unless the data fitting interval is very large. For larger number of modes the high frequencies are not well estimated, and generate large oscillations of the fitting function beyond the data assimilation time.
 \item
 To assimilate data for the true pandemic, a time adjustment of the basic reproductive rate appears to model the relaxation in the fulfillment of the lockdown measures.
 \end{itemize}
 The application of the procedure to the evolution of the Covid-19 pandemic in Andalusia region, Italy and Spain shows accurate predictions for 5 days, that improve as the number of assimilated data increases.


\begin{thebibliography}{00}
 \bibitem{Andaludat} https://www.juntadeandalucia.es/institutodeestadisticaycartografia/badea/operaciones/ consulta/anual/38228?CodOper=b3\_2314\&codConsulta=38228.
 
 \bibitem{Italydat}  https://github.com/pcm-dpc/COVID-19/blob/master/dati-andamento-nazionale/dpc-covid19-ita-andamento-nazionale.csv.
 
 \bibitem{Spaindat}  https://covid19.isciii.es/resources/serie\_historica\_acumulados.csv.
 
\bibitem{RPOD} {\sc Mejdi Aza\"iez, Faker Ben Belgacem,  Tom\'as Chac\'on Rebollo}, {\it Recursive POD expansion for reaction-diffusion equation} Eng. Sci. 3:3 (2016), DOI 10.1186/s40323-016-0060-1

\bibitem{Varona} {\sc Jos\'e M. Guti\'errez, Juan L. Varona},
{\it An\'alisis de la posible evoluci\'on
de la epidemia de coronavirus COVID-19
por medio de un modelo SEIR}, https://www.unirioja.es/apnoticias/servlet/Archivo?C\_BINARIO=12051 (2020)

\bibitem{chinos} {\sc Jia Jiwei, Ding Jian, Liu Siyu, Liao Guidong, Li Jingzhi, Duan Ben, Wang Guoqing, Zhang Ran},
{\it Modeling the Control of COVID-19: Impact of Policy Interventions and Meteorological Factors}, arXiv:2003.02985v1 (2020).


\bibitem{muller} {\sc M. Muller M.}, {\it On the POD Method. An Abstract Investigation with Applications to Reduced-Order Modeling and Suboptimal
Control}.  Ph D Thesis. Georg-August Universit\"at, G\"ottingen (2008).
 
\bibitem{franceses} {\sc Lionel Roques, Etienne Klein, Julien Papa\"{\i}x, Samuel Soubeyrand},
{\it Mod\`ele SIR m\'ecanico-statistique pour l'estimation du nombre d'infect\'es et du taux de mortalit\'e par COVID-19}, arXiv:2003.10720v2 (2020).


%
\bibitem{Muller} {\sc M. M\"uller,} On the POD Method. An Abstract Investigation with Applications to Reduced-Order Modeling and Suboptimal
Control, Ph D Thesis. Georg-August UniversitŠt, Gšttingen; 2008.
\end{thebibliography}
\end{document}